\documentclass[fleqn,usenatbib]{mnras}

\usepackage{newtxtext,newtxmath}
\usepackage[T1]{fontenc}

\DeclareRobustCommand{\VAN}[3]{#2}
\let\VANthebibliography\thebibliography
\def\thebibliography{\DeclareRobustCommand{\VAN}[3]{##3}\VANthebibliography}

\usepackage{graphicx}	% Including figure files
\usepackage{amsmath}	% Advanced maths commands
\usepackage{xspace}

\newcommand{\orcid}[1]{\href{http://orcid.org/#1}{\includegraphics[width=9pt]{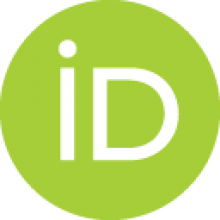}}}

\newcommand{\Rfreq}{\ensuremath{R_{\text{Freq}}}\xspace}
\newcommand{\RILR}{\ensuremath{R_{\text{ILR}}}\xspace}
\newcommand{\RILRw}{\ensuremath{R^{\text{W}}_{\text{ILR}}}\xspace}
\newcommand{\RILRh}{\ensuremath{R^{\text{H}}_{\text{ILR}}}\xspace}
\newcommand{\RILRc}{\ensuremath{R^{\text{C}}_{\text{ILR}}}\xspace}
\newcommand{\thetabar}{\ensuremath{\theta_{\text{bar}}\xspace}}

\defcitealias{lucey2023}{L23}
\defcitealias{beraldoesilva2023}{BS23}
\defcitealias{hey2023}{H23}
\title[Dynamical modelling of the bar]{Self-consistent dynamical modelling of the Milky Way bar with orbital frequency analysis}

\author[Z. Langford et al.]{
Zachary Langford \orcid{0000-0001-7574-4440}$^{1}$\thanks{E-mail: langfzac@sas.upenn.edu},
Robyn Sanderson \orcid{0000-0003-3939-3297}$^{1}$,
Madeline Lucey \orcid{0000-0001-7297-8508}$^{1}$,
and Jason A. S. Hunt \orcid{0000-0001-8917-1532}$^{2}$
\\
$^{1}$Department of Physics and Astronomy, University of Pennsylvania, Philadelphia, PA 19104, USA\\
$^{2}$School of Mathematics and Physics, University of Surrey, Guildford, Surrey GU2 7XH, UK\\
}

\date{Accepted XXX. Received YYY; in original form ZZZ}

\pubyear{2026}

\begin{document}
\label{firstpage}
\pagerange{\pageref{firstpage}--\pageref{lastpage}}
\maketitle

\begin{abstract}
We present an update to the frequency analysis method for measuring the properties of a galactic bar. The method involves computing the fundamental frequencies of orbits in rotating, N-body-derived potential models, classifying the stars as members of bar supporting orbits, and finding the extent of the apo-centre distribution. In this work, we apply an updated classification criterion designed to isolate the so-called ``Warm" inner Lindblad resonance (ILR) orbits. These orbits have been shown to contain the looped x1 orbits, which dominate the ``shoulder regions" of the bar and largely contribute to the radial extent. We apply this method to existing Gaia, APOGEE, and OGLE data of more than 200,000 stars to constrain the properties of the Milky Way bar. We find that multiple bar lengths and pattern speeds are consistent with the data to within 5 percent.
\end{abstract}

\begin{keywords}
Galaxy: structure -- galaxies: bar -- galaxies: kinematics and dynamics -- methods: data analysis -- methods: numerical
\end{keywords}

%%%%%%%%%%%%%%%%%%%%%%%%%%%%%%%%%%%%%%%%%%%%%%%%%%

%%%%%%%%%%%%%%%%% BODY OF PAPER %%%%%%%%%%%%%%%%%%

\section{Introduction}
The inner regions of many disk galaxies have been observed to host stellar bars \citep{Sellwood1993,Masters2011,Gavazzi2015}. These bars appear as extended over-densities in the surface brightness profile that seem to rotate as a solid body with morphologies that depend on the particular resonant orbits of which they are comprised \citep[e.g.,][]{Petersen2021}. Galactic bars drive a number of resonances within the disk \citep{Dehnen2000,Minchev2010,Antoja2018,Hunt2018,Fujii2019}, and contribute substantially to the overall potential in the inner regions of the host galaxy \citep[e.g.,][]{Portail2015}.

Understanding the morphological properties of a bar is one way towards understanding it's potential and distribution function. The pattern speed determines the possible resonant orbit families that can exist in a given potential, and is therefore linked to the bar's strength and length \citep{athanassoula1992}. However, the pattern speed does not strictly determine the length, as the available resonant orbit families may or may not be populated for a given galaxy. Several metrics have been used to measure the length, angle, and pattern speed of a bar, for both simulated and observed galaxies. This includes kinematic studies \citep[e.g.,][]{termaine1984,Petersen2024} or computing the strength of Fourier modes in the surface density profile \citep[e.g.,][]{rosasguevara2020}. In simulations, the angle and pattern speed can be well constrained using the latter. Since the bar rotates as if it were a solid body, the effects of perturbations near the edges of the bar (e.g. spiral arms) on these measurements can easily be mitigated. However, those same edge effects complicate the measurement of the bar's length \citep{Gonzalez2016,Hilmi2020}. 

\citet[][hereafter \citetalias{lucey2023}]{lucey2023} describe a method to measure the ``dynamical length" of a galactic bar via fundamental frequency analysis. The dynamical length is the maximal extent of the apo-centre distribution of the bar-resonant orbits. We may determine a given orbit's membership to the bar by computing the fundamental frequencies from a numerically integrated trajectory in a model potential. The authors show that a set of test particles will only replicate the intrinsic dynamical length of the underlying model potential if the test particle set is a representative sample of the underlying distribution function. That is, the potential model must have a similar bar length and pattern speed to that of the test particle distribution. In addition to the measurement of the bar length, this method also provides a constraint (by way of a specific model) on the pattern speed of the bar and it's underlying potential.

This method, as well as the other length measurement methods, depends on a few choices. Namely, which population of orbits to target. \citetalias{lucey2023} point out that their selection method misses the $x_1$ family of orbits, which is known to be a major component of galactic bars \citep{wang2016}, and are known to contribute greatly to the radial extent \citep[e.g.][]{beraldoesilva2023}. The $x_1$ family of orbits are loop orbits that predominantly lie along the bar's major axis (in the bar-rotating frame) and surround the inner-Lindblad resonance(s) \citep{contopoulos1977,contopoulos1980}. \citet[][hereafter \citetalias{beraldoesilva2023}]{beraldoesilva2023} show that the bar ``shoulder" regions that contain the looped $x_1$ orbits can be effectively isolated by using a different cut in frequency space. Further, their method enables finer control over which bar resonant stars are included in the analysis. The additional cuts in frequency space allow us to target stars with orbits that are more traditional ``bar orbits" and exclude those that are still resonant orbits, but clearly do not contribute to the bar morphology.

\begin{figure*}
    \centering
    \includegraphics[width=\textwidth]{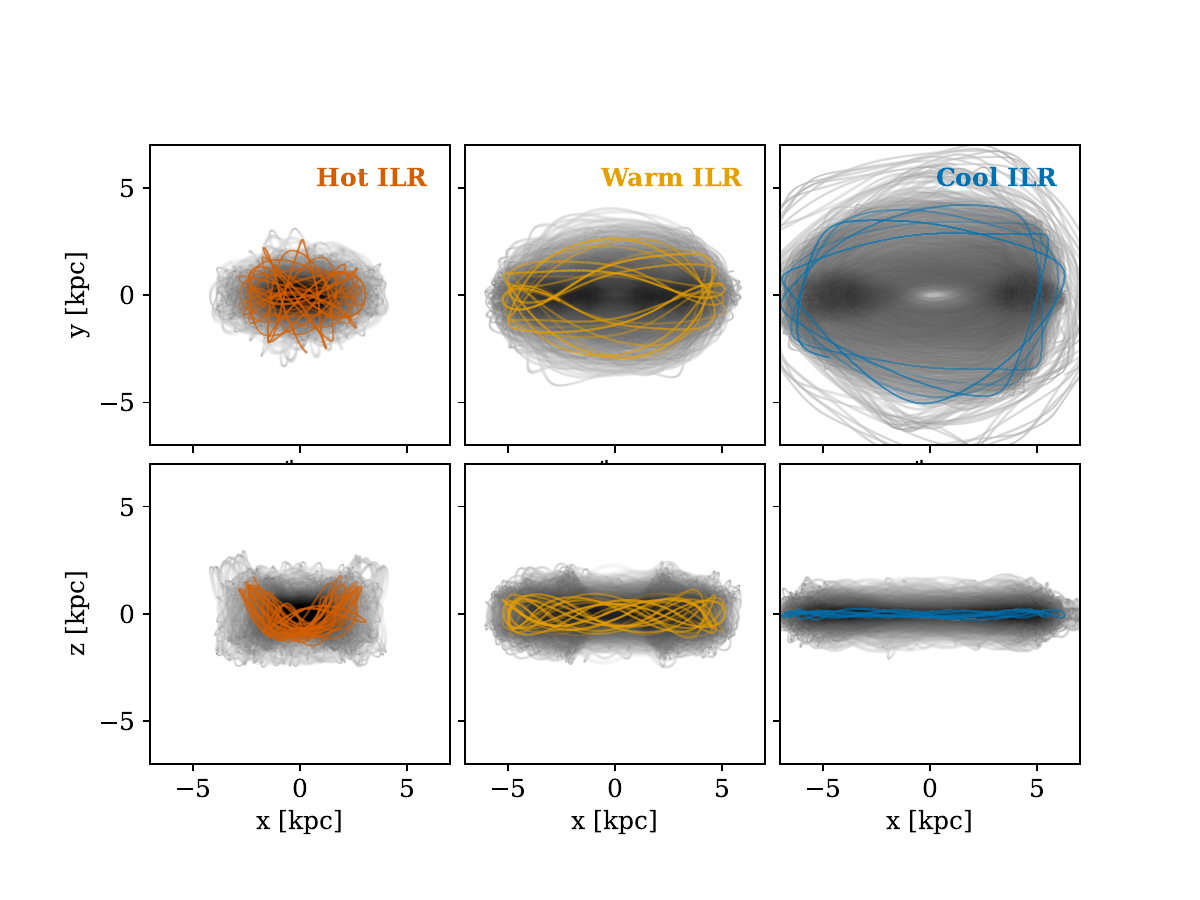}
    \caption{Selection of orbits for each ILR population (coloured lines) and the projected, stacked orbits of each of the sampled N-body particles (gray density; all on the same scale). The hot ILR are in red (left), warm ILR are in orange (middle), and cool ILR are in blue (right).}
    \label{fig:orbits}
\end{figure*}

The authors of \citetalias{lucey2023} apply their method to the MW by integrating stars from Gaia eDR3 \citep{gaia,edr3} in MW-like potential models, with a variety of bar lengths and pattern speeds, and comparing the resulting dynamical lengths to the intrinsic dynamical lengths of the model potentials. The distances used in \citetalias{lucey2023} are given by \texttt{astroNN} \citep{astronn,astronn_distances} -- a deep neural-network model trained on overlapping distances in Gaia and APOGEE \citep{apogee} data sets. Along with the authors spatial cuts, the catalogue provides 6D coordinates for over 200,000 stars between the sun and the galactic centre. Recently, \cite{hey2023} have released a new distance catalogue for stars in the MW bulge using the Optical Gravitational Lensing Experiment \citep[OGLE][]{soszynski2004,soszynski2009}. Combining this with existing Gaia data we obtain 6D phase-space coordinates for an additional 40,413 stars, many extending to the far-side of the Galactic bulge.

In this paper, we employ an updated frequency analysis method to measure the properties of the MW bar. We use the frequency analysis method from \citetalias{lucey2023}, but with the bar-star selection criterion from \citetalias{beraldoesilva2023} (Section \ref{sec:measure_bar_length}). This updated method is able to better determine which stars are members of the bar, including the so-called ``shoulder-region", which corresponds more closely to a morphological description. We use the same N-body simulation as in \citetalias{lucey2023} (Section \ref{sec:nbody_model}), from which we derive Basis Function Expansion (Section \ref{ssec:pot_models}) models for orbit integration. We use the same catalogue of stars as in \citetalias{lucey2023} and incorporate an additional 40,413 stars that have distance measurements derived from the OGLE survey (Section \ref{sec:data}). We describe our results and discuss the properties of our ``best fit" potential models in Section \ref{sec:results}. The data sets used in this work are already publicly available, the N-body snapshots are publicly available\footnote{\url{https://users.flatironinstitute.org/~jhunt/Bennett_MWP14-3_MR_snapshots/}}, and all of the code used for the analysis and figures for this paper will be available in a public GitHub repository\footnote{\url{https://github.com/langfzac/MW_bar_paper}}.

{\subsection{A note on colour}
Throughout this paper, our aim is to use an intuitive colour scheme to distinguish between different populations of orbits. We have chosen to use the accessible \cite{wong2011} colour palette for discrete points. These colours differ slightly from the usual red, blue, green, and so on. For brevity (and clarity), the \cite{wong2011} colours ``bluish green", ``Vermillion", and ``reddish purple" will be referred to as simply green, red, and purple.

\section{Measuring the length of a bar}\label{sec:measure_bar_length}
In this section, we outline our method for measuring the dynamical length of a galactic bar. We differentiate this length from other length measurements, such as the Fourier length \citep[e.g.,][]{rosasguevara2020}, in that these techniques rely specifically on the dynamics of resonant stars to measure the bar morphology. Throughout this work, we will refer to any estimate of the radial extent of bar-resonant orbits as a dynamical length and any specific estimators are defined where necessary.

We follow the general framework outlined in \citetalias{lucey2023}, which describes measuring the dynamical length of a barred potential model using fundamental frequency analysis of the orbits of test particles. The particular frequencies assigned to each orbit mark which stars are members of the bar supporting orbit families. We estimate the dynamical length using the same 99.5-th percentile of the apo-centre distance distribution as in \citetalias{lucey2023}. This is designed to measure the radial extent of bar supporting orbits while removing any misclassified large apo-centre outliers. To be consistent with \citetalias{lucey2023}, we define an estimate of the dynamical length of a bar using their specific frequency cuts to be \Rfreq.

In principle, the dynamical length is an intrinsic property of a given potential (with a set pattern speed). However, any measurement of it also depends on the set of test-particles we integrate -- the available orbit families need not be populated. So, for a given potential, any estimate of the dynamical length (e.g., \Rfreq) is only the dynamical length \textit{given} that specific set of orbits. To estimate the ``intrinsic" dynamical length of a specific potential, we require a representative sample from that potentials underlying distribution function. \citetalias{lucey2023} show that this is also true the other way. That is, given a set of test-particles, one can ``fit" a potential model to them by comparing the \Rfreq they produce in a given potential model to the potential models intrinsic \Rfreq.

In order to measure the dynamical length of the MW, \citetalias{lucey2023} compare the intrinsic \Rfreq of MW-like N-body simulation snapshots to the \Rfreq derived from the dynamics of Gaia eDR3 stars integrated in the same potential. The authors start by fitting a basis-function expansion model (BFE) to each N-body simulation snapshot and sample initial conditions from the N-body particles. They then integrate the orbits in the BFE potential with a fixed pattern speed, measure the fundamental frequencies, and determine the extent of the bar stars. This defines the intrinsic \Rfreq for a given BFE potential model and pattern speed. Next, they take 6D phase-space data for real stars in the Gaia eDR3 catalogue as initial conditions, integrate them in the same potential model, and compare the resulting \Rfreq to the potentials intrinsic \Rfreq. The ``best-fit" model (i.e. the \Rfreq that is most consistent with the potential model's intrinsic \Rfreq) is taken to be the measurement of the MW bar dynamical length. 

In place of the selection used in \citetalias{lucey2023}, we use the frequency space cuts outlined in \citetalias{beraldoesilva2023}. The authors show that stars near the $\left(\Omega_{\phi} - \Omega_P\right)/\Omega_R = 0.5$ resonance are an important marker for the radial extent of the bar -- the so-called ``shoulders". Here, $\Omega_P$ is the pattern speed of the rotating potential, and $\Omega_{R}$ and $\Omega_{\phi}$ are the cylindrical galactocentric coordinate frequencies in the inertial frame. The resonance is a more general interpretation of the inner-Lindblad resonance (ILR) \citep[e.g.,][]{athanassoula2003}, and the authors split up the resonance into three populations: Hot, Warm, and Cool ILR. Each population is a different cut in the vertical frequency, which is a proxy for vertical extent. We choose to isolate the warm ILR population, which contains the looped $x_1$ orbits at the shoulders. The Hot ILR orbits at $\Omega_z/\Omega_R < 1$ are quite vertically extended, have little consistency in shape, and lie much closer to the galactic centre. The Cool ILR orbits at $\Omega_z/\Omega_R > 3/2$ are much more vertically flattened and elliptical, and while they do contribute to the extent of the shoulders, they also contain orbits that extend much further and bias the resulting length measurement. Figure \ref{fig:orbits} shows example orbits for each of these populations; the details of the figure are described in Section \ref{ssec:nbody_bars}. For dynamical length estimates made using this updated set of cuts and the same 99.5-th percentile in apo-centre distance, we use \RILRh, \RILRw, and \RILRc for the Hot, Warm, and Cool ILR lengths, respectively. When the specific ILR population is irrelevant, we will simply use \RILR to refer to a generic ILR dynamical length estimate.

\subsection{Frequency Analysis}\label{ssec:freq_analysis}
Numerical Analysis of Fundamental Frequencies (NAFF) for orbits in galaxy potentials is well described and applied in other works \citep[e.g.][]{Laskar1992,valluri2016}. Here, we aim to provide a brief explanation of the technique and the particular implementation we used.

We may approximate the time series of a phase-space coordinate using a finite Fourier series:
\begin{equation}
    f(t) = \sum_{n=1}^{N} c_n \exp(i\omega_n t)
\end{equation}
where $c_n$ is the usual complex Fourier coefficient, $\omega_n$ is the frequency of the $n$-th term, and $N$ is the number of terms in the approximation.

We use the Python package \texttt{naif} \citepalias{beraldoesilva2023} to find the largest Fourier coefficient, which corresponds to the fundamental (leading) frequency\footnote{This leading frequency is not guaranteed to be the fundamental frequency, but is usually the case.}. The coefficients are found via the usual inner product of the windowed time-series and the relevant exponential term in the Fourier series: 
\begin{equation}
\begin{split}
    c_n & = \left<f(t)W(t), \exp(i\omega_n t) \right> \\
        & = \frac{1}{T} \int_{-T/2}^{T/2}f(t) W(t) \exp(-i\omega_n t) \,dt
\end{split}
\end{equation}
where $W(t)$ is a window function\footnote{\texttt{naif} uses the Hanning window, and we choose $p=1$. See Appendix A and Equation A2 in \citetalias{beraldoesilva2023}.} and $T$ is the total time interval spanned by $f(t)$. To find the peak, the code identifies an estimate from the frequency spectrum, computed using a windowed, discrete Fourier transform. Then, the estimate is refined by optimizing the inner product over frequency.

\subsection{Orbit integration}
For the numerical orbit integration, we use the open-source Python/C++ package \texttt{agama} and their implementation of an 8th order Runge-Kutta integrator with adaptive time-stepping \citep{agama}. We compute the orbits in single-float precision (\texttt{agama} default), which cuts down computation time and memory/storage requirements, while not meaningfully affecting the results of the frequency analysis. We carry out the integration using a rotating potential in the inertial frame.

\section{N-body Model}\label{sec:nbody_model}
The simulation used in this work is the $N$-body simulation `Galaxy A' from \citetalias{lucey2023}. This model was itself a reproduction of the MWP14-3 Milky Way-like host galaxy from \cite{bennett2022}. The simulation parameters are derived from a version of \texttt{Galpy}\rm's \texttt{MWPotential2014}\rm\ \citep{bovy2015}, with a more massive dark matter halo ($M_{\mathrm{200}}=1.4\times10^{12}\ M_{\odot}$). The full process of setting up the initial conditions, both in terms of parameter selection and the conversion from the formalism of \texttt{Galpy}\rm\ to that of the initial condition generator \texttt{GalIC}\rm\ \citep{yurin2014} is described in \cite{bennett2022}. 

The `Galaxy A' simulation from \citetalias{lucey2023} is then a reconstruction of this model with $\sim3.3\times10^6$ particles in the dark halo, $5\times10^6$ particles in the stellar disc, and $\sim1\times 10^6$ particles in the stellar bulge. The initial condition was generated with \texttt{GalIC}\rm, and the simulation was evolved in isolation for 5 Gyr with the GPU accelerated $N$-body tree code \texttt{Bonsai}\rm\ \citep{bedorf2012,bedorf2014}. 

\subsection{Potential Models}\label{ssec:pot_models}
We use \texttt{agama} to fit a basis-function expansion (BFE) model to each N-body snapshot. Each particle population is fit with a distinct model. The dark matter halo and bulge stars are modelled with `Multipole' expansions made up of a sum of spherical-harmonic angular functions times a radial quintic spline. We choose a radial grid size of 20 and an $l_{max}$ of 6 for the angular functions. The stellar disk is modelled using a `CylSpline', which is a 2D quintic spline for the radial and z dimensions and a Fourier term for the angular dependence. We choose a grid size of 20 for the radial and z splines, restrict the interpolation to $R < 60.0$ kpc and $|z| < 10.0$ kpc, and fix the order of the angular function to an $m_{max}$ of 6. We refer the reader to the \texttt{agama} documentation \citep{agama_doc} for details of each BFE implementation.

Once we have a BFE for each snapshot, we measure the bar angle and pattern-speed using the Fourier decomposition of the surface density of the stellar components. We find the $m$-th Fourier mode of the $i$-th snapshot via the inner-product: 

\begin{equation}
    A_{m,i}(r) = \frac{1}{\pi}\int_{0}^{2\pi} \Sigma\left(r, \theta\right) \cos\left(m\theta\right) d\theta
\end{equation}
and 
\begin{equation}
    B_{m,i}(r) = \frac{1}{\pi}\int_{0}^{2\pi} \Sigma\left(r, \theta\right) \sin\left(m\theta\right) d\theta
\end{equation}
where $r$ is the radius from the centre, $\theta$ is the azimuthal angle, and $\Sigma_i\left(r, \theta\right)$ is the surface density where $z$ = 0. In practice, these integrals are computed numerically using \texttt{scipy.quad} \citep{scipy}.

We measure the bar angle for each snapshot via
\begin{equation}
    \thetabar(t_i, r) = \frac{1}{2}\arctan{\left(\frac{B_{2,i}(r)}{A_{2,i}(r)}\right)}.
\end{equation}
where $t_i$ is the time of the $i$-th snapshot. We compute \thetabar$(t_i, r)$ over 10 values of R, evenly spaced from 1.5 to 3 kpc, and take the median to be \thetabar($t_i$).

To measure the pattern-speed, $\Omega_p$, we use forward finite-differences to compute the time-derivative of the bar angle for a given snapshot.
\begin{equation}
    \Omega_p(t_i) = \frac{\thetabar(t_{i+1}) - \thetabar(t_i)}{t_{i+1} - t_i}
\end{equation} 

\subsection{Sample Selection}\label{ssec:sample_selection}
To measure the intrinsic dynamical lengths of the N-body snapshots, we must integrate test particles in the corresponding BFE potential. From a subset of the N-body snapshots, we take a random sample of 10,000 stars within a cylindrical volume defined by $R < 10.0$ and $|z| < 3.0$. To choose the subset of snapshots, we model the overall evolution of the pattern speed by a simple $A/\Omega_p + B$ functional form. We fit the parameters A and B using least-squares, and only include pattern speed measurements for $t > 2.5$ Gyr. We choose the pattern speed measurements that are within 1.5 km/s/kpc from the best-fit model, and thin the sample such that we have a spacing of about 100 Myr between snapshots. 

Figure \ref{fig:pattern_speeds_samples} shows the measured pattern-speeds of each N-body snapshot (blue, dots) for times greater than 2 Gyr along with the best fit model (black, dashed line) and the selected snapshots (red, dots).

\subsection{N-body Properties}\label{ssec:nbody_bars}
Table \ref{tab:snap_samples} shows the properties of each selected potential model, including the simulation time of the snapshot, the bar angle from the positive x-axis, and the instrinsic \RILR measured from our frequency analysis. We name the potential models after the corresponding snapshot as \texttt{agama\_00XXX} (which matches the file name in the provided source code for this work).

Figure \ref{fig:orbits} shows example orbits for the \texttt{agama\_00419} model. The coloured orbits are selected to represent a typical orbit near the outer edge of the apo-centre distribution of each population\footnote{For other examples of these orbits, see Figure 7 in \citet{beraldoesilva2023}}. The Hot ILR are in red (left), the Warm ILR are in orange (middle), and the Cool ILR are in blue (right). The under-plotted density map represents the projected, stacked orbits of each of the sampled N-body particles that are members of the labelled ILR population. 

\begin{figure}
    \centering
    \includegraphics[width=\linewidth]{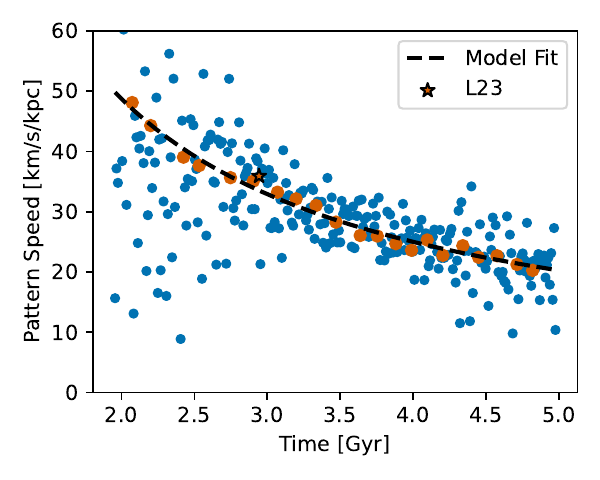}
    \caption{The pattern speed [km/s/kpc] of an N-body snapshot versus simulation time [Gyr]. The blue points are all available N-body snapshots. The black, dashed line is the model fit to the blue points via least-squares. The larger red points are a selection of snapshots that are near the fit and spaced quasi-evenly in time (Section \ref{ssec:sample_selection}). We also include the ``best-fit" model from \citetalias{lucey2023} (\texttt{agama\_00301}) as the star-shaped point.}
    \label{fig:pattern_speeds_samples}
\end{figure}

\begin{table}
\centering
\caption{Properties of the selected N-body potential models (see Figure \ref{fig:pattern_speeds_samples}). The snapshot number; the simulation time of the snapshot (Gyr); the measured pattern-speed of the snapshot (km/s/kpc); the angle of the bar from the positive x-axis in degrees; the Warm ILR dynamical length of the bar (kpc).}
\begin{tabular}{cccccc}
    \hline
    Snapshot & Time & Pattern Speed & Angle & \RILRw \\
    & Gyr & km/s/kpc & deg & kpc \\ 
    \hline
    212 & 2.073 & 48.091 & -62.437 & 1.564 \\
    225 & 2.2 & 44.274 & 34.818 & 1.703 \\
    248 & 2.425 & 39.037 & 1.35 & 2.305 \\
    259 & 2.533 & 37.638 & 49.715 & 2.643 \\
    281 & 2.748 & 35.639 & -20.837 & 3.305 \\
    297 & 2.904 & 35.07 & -57.684 & 3.404 \\
    301 & 2.943 & 35.901 & 24.084 & 3.242 \\
    314 & 3.07 & 33.228 & -88.585 & 3.789 \\
    327 & 3.197 & 32.122 & -31.462 & 4.114 \\
    341 & 3.334 & 31.027 & 39.696 & 4.447 \\
    355 & 3.471 & 28.26 & -81.467 & 4.697 \\
    372 & 3.637 & 26.051 & 13.296 & 5.193 \\
    384 & 3.755 & 25.952 & 25.964 & 5.522 \\
    397 & 3.882 & 24.69 & 39.945 & 5.661 \\
    408 & 3.989 & 23.595 & 22.32 & 5.844 \\
    419 & 4.097 & 25.305 & -2.405 & 5.94 \\
    430 & 4.205 & 22.755 & -27.805 & 6.32 \\
    444 & 4.342 & 24.353 & -17.176 & 5.887 \\
    455 & 4.449 & 22.439 & -56.792 & 6.182 \\
    468 & 4.576 & 22.691 & -69.0 & 6.245 \\
    482 & 4.713 & 21.241 & -79.777 & 6.691 \\
    493 & 4.821 & 20.272 & 52.862 & 6.849 \\
    \hline
\end{tabular}
\label{tab:snap_samples}
\end{table}

Figures \ref{fig:freq_cuts_samples} and \ref{fig:freq_z_apo_samples} show the fundamental frequencies for orbits of the selected N-body particles in the \texttt{agama\_00419} potential model. The left panel are rotating frame Cartesian frequency ratios and the right panel shows the same stars in inertial frame cylindrical coordinate frequencies. 

In Figure \ref{fig:freq_cuts_samples}, the coloured points correspond to different orbit populations: purple are the \citetalias{lucey2023} Bar cuts, red are Hot ILR, orange are Warm ILR, blue are Cool ILR, and green are the corotation resonant stars. We note that all of the ILR stars that overlap with the purple-shaded region are also found by the L23 method. In order to illustrate the difference between methods, we show the same stars in both panels.

In Figure \ref{fig:freq_z_apo_samples}, the colours represent the apo-centre (maximum cylindrical radius, R) distance of the orbit (top panels) and the maximum z distance (bottom panels) for the \citetalias{lucey2023} Cartesian frequency cuts and the Hot, Warm, and Cool ILR stars. The left and right panels match the coordinates in Figure \ref{fig:freq_cuts_samples}. The outlined regions in the left panels of Figure \ref{fig:freq_z_apo_samples} are a visual aid to show roughly where the bulk of ILR stars overlap with the Cartesian frequency cuts. The other ILR stars above and below this region lie along distinct lines and can be more easily distinguished visually. 

These figures further motivate our choice to use cylindrical frequency cuts. The Cartesian cut misses many of the ILR stars (Fig. \ref{fig:freq_cuts_samples}), while including high apo-centre orbits of many non-ILR bar resonant stars (Fig. \ref{fig:freq_z_apo_samples}).

\begin{figure*}
\centering
\includegraphics[width=\linewidth]{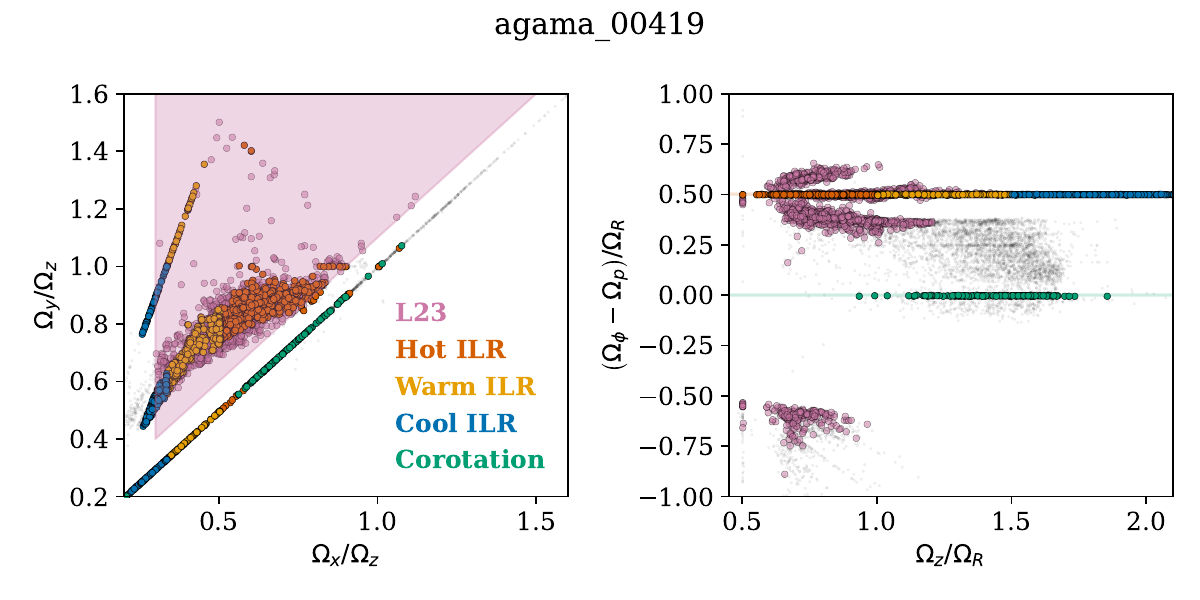}
    \caption{Fundamental frequencies for orbits of N-body particles in their corresponding potential model (\texttt{agama\_00419}). The red points represent the hot ILR, the orange are warm ILR, the blue are cool ILR, the green are corotation resonance stars, and the black points are any stars not classified into any of the preceding populations. Left panel: Cartesian frequency ratios in the bar-rotating frame. The purple shaded region is the cut used in \citetalias{lucey2023}. Right panel: Cylindrical coordinate frequencies in the inertial frame. The purple points are those that lie in the shaded region in the left panel. The \citetalias{lucey2023} cut misses many ILR stars while including non-ILR stars.}
    \label{fig:freq_cuts_samples}
\end{figure*}

\begin{figure*}
    \centering
    \includegraphics[width=\linewidth]{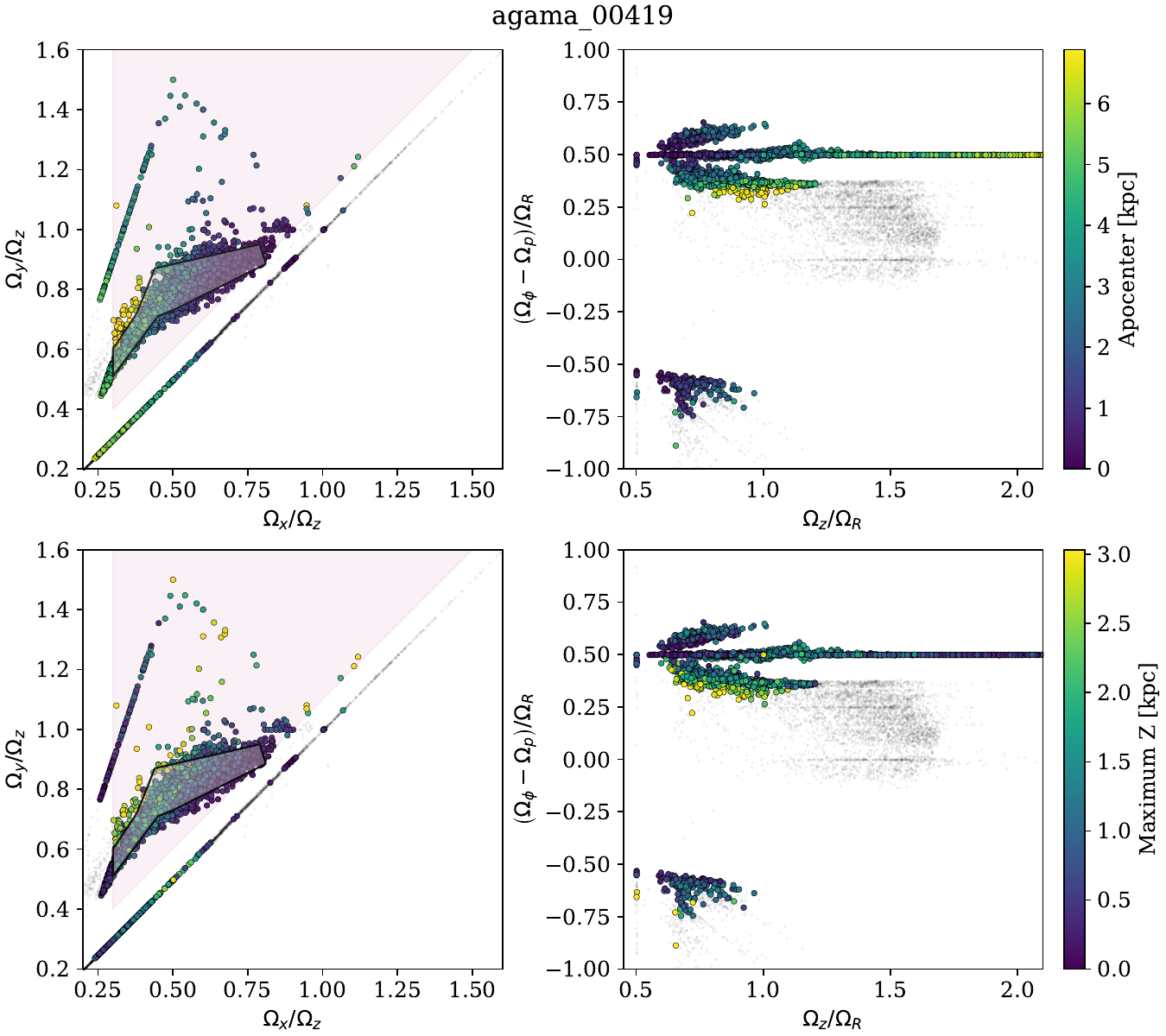}
    \caption{Fundamental frequencies for orbits of N-body particles in their corresponding BFE potential (\texttt{agama\_00419}; same as Figure \ref{fig:freq_cuts_samples}). The left and right panels are the same as in Figure \ref{fig:freq_cuts_samples}, the colours now represent (top) the apo-centre distance and (bottom) the maximum z extent along the orbit for the \citetalias{lucey2023} cuts and the Hot, Warm, and Cool ILR stars. The outlined region in the left panels are a visual aid to show roughly where the bulk of ILR stars overlap the Cartesian frequency cuts in Figure \ref{fig:freq_cuts_samples}. We see that choosing ILR stars to measure the dynamical length removes a region of large apo-centre distances.} 
    \label{fig:freq_z_apo_samples}
\end{figure*}

\section{Data}\label{sec:data}
In this section, we describe the data sets used as initial conditions for orbit integration. Figure \ref{fig:data} shows the full data selection, transformed to galactocentric coordinates (Section \ref{ssec:data-6d}). The colour-bar represents the number of stars from the astroNN catalog (Section \ref{ssec:astronn}), and the orange points\footnote{The ``banding" in the orange data points is due to the transformation from distance modulus to physical distance. The precision of the input catalogue is to one decimal place, so going from a distance modulus of, say, 15 to that of 15.1 corresponds to a 471 pc difference in physical distance.} are from the OGLE-III survey (Section \ref{ssec:ogle}). 

\subsection{astroNN}\label{ssec:astronn}
We use a subset of the astroNN \citep{astronn,astronn_distances} value-added catalogue for this analysis. The data is produced from additional processing of the APOGEE DR17 \citep{apogee,dr17} and Gaia eDR3 \citep{gaia,edr3} data sets. Gaia provides precise RA, DEC, and proper-motion measurements, but the reported parallax distances to stars further than a few kpc from the sun can be unreliable. Spectroscopic surveys like APOGEE are able to measure distances beyond this to high precision. The astroNN model uses a deep neural network, trained on overlapping Gaia and APOGEE data to determine distances to stars in Gaia.

In order to propagate uncertainties to our measurement of the bar length, we use the measurement errors and correlation coefficients for RA, DEC, and proper motions from Gaia eDR3, and the radial velocity errors from astroNN/APOGEE. 

To summarize, the RA, DEC, proper motions, and the associated errors and correlation coefficients are taken from Gaia eDR3, the distance and error are from astroNN, and the radial velocity and error are from APOGEE DR17. The astroNN catalog includes Gaia IDs, which we use to obtain the relevant Gaia eDR3 data. The radial velocity measurements are given in the astroNN dataset, but the uncertainties must be pulled from APOGEE, using the provided APOGEE IDs. To be consistent with \citetalias{lucey2023}, we do not make any additional quality cuts outside of filtering missing values and NaNs.

\subsection{OGLE}\label{ssec:ogle}
In addition to astroNN, we utilize the catalogue of distances to stars near the galactic centre from \citet[][hereafter \citetalias{hey2023}]{hey2023}. The authors measure distance moduli to OGLE-III semi-regular variables (evolved red giants) \citep{OGLEIII} using a period-amplitude-luminosity relation. We convert these measurements to distance via the usual distance modulus equation:
\begin{equation}
    \log_{10}(d) = 1 + \frac{\mu}{5}
\end{equation}
where $\mu$ is the distance modulus, and $d$ is the distance in parsecs. We then propagate uncertainties using the Monte-Carlo method. Each measurement and uncertainty is taken to be the mean and standard deviation of a uni-variate Gaussian distribution, from which we draw 100,000 samples. We then use the median and standard deviation of the samples as the distance and associated uncertainty. 

To obtain the full set of celestial coordinates, we cross-match \citetalias{hey2023} with Gaia DR3 to 1 arc-second. We make additional cuts for RUWE $<$ 1.3, fractional radial velocity error $<$ 25, and fractional distance errors $<$ 25\%. 

Our signal-to-noise selection implicitly throws out stars with small radial velocities, which could potentially bias a bar length measurement since stars at apo-centre have intrinsically low radial velocities. However, we have verified that the position and tangential velocity distributions of the eliminated stars are consistent with the overall sample distribution. Therefore, we do not expect this selection to contribute any meaningful kinematic bias.

\subsection{6D Phase-space Coordinates}\label{ssec:data-6d}
Using the astropy v4.0 definition, we convert the catalogue data to galactocentric coordinates, and propagate errors using the Monte-Carlo method. We use the full covariance matrix (constructed from the errors and correlation coefficients) of the data to define, and take 10,000 samples from, a multivariate Gaussian distribution for each object. We then compute the mean, standard deviation, and correlation coefficients from the set of samples to use as positions, velocities, and associated uncertainties.

For the astroNN stars, we make the spatial cuts $|\text{Y}|$ $<$ 10 kpc and -8.0 $<$ $\text{X}$ $<$ 0 kpc, where the sun is located at -8.122 kpc along the X-axis. The final set of astroNN objects contains 198,242 stars.

For OGLE, we make the spatial cuts $|\text{Y}|$ $<$ 10 kpc and $\left|\text{X}\right| < 8.0$ kpc. We increase the spatial extent for OGLE, as we expect the distance uncertainties to be reliable in this range. These cuts leave us with 40,413 stars.  

Finally, we rotate the data 180 degrees about the y-axis such that the sun would lie along the positive x-direction and the bulk rotation would be counter-clockwise -- aligning with the rotation of the N-body simulation. 

\begin{figure*}
    \centering
    \includegraphics[width=\linewidth]{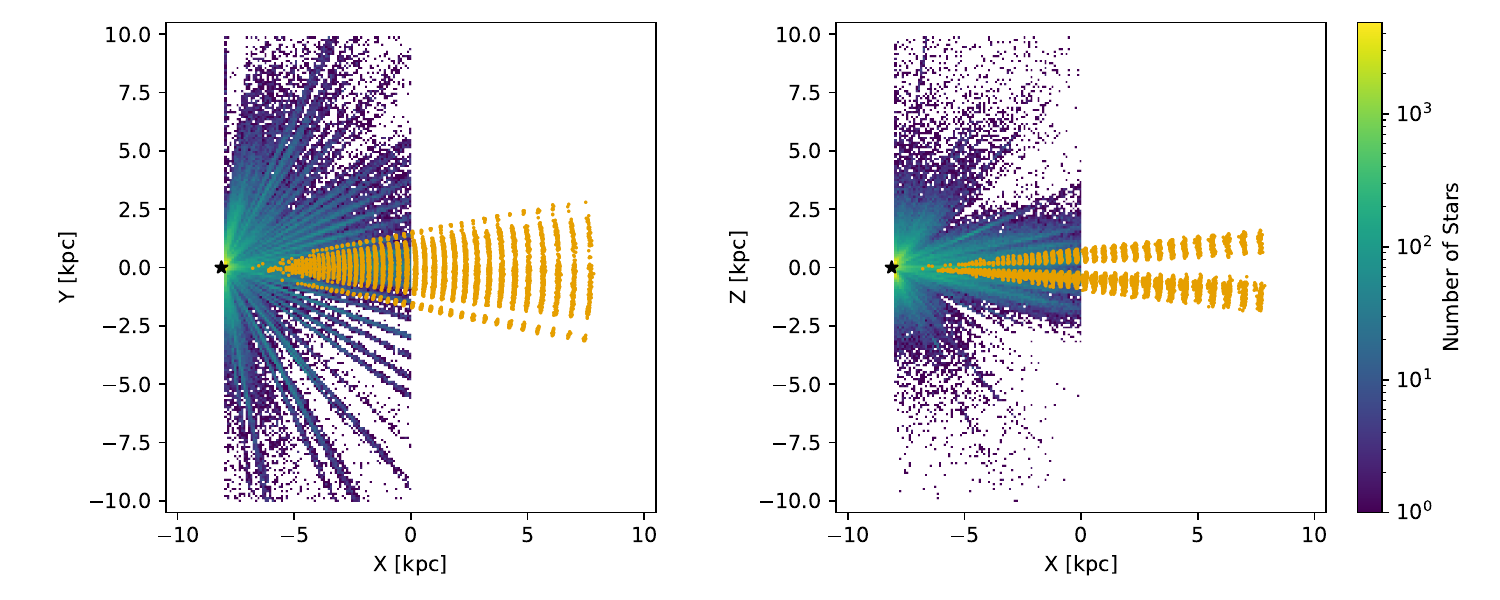}
    \caption{Representation of the astroNN and OGLE data in Galactocentric Cartesian coordinates. The left panel displays the x-y coordinates in kpc, and the right panel shows the x-z coordinates. The colour-bar represents the number of stars in a given spatial bin for the astroNN stars. The orange points are individual stars from the OGLE data set. This figure include all the spatial cuts outlined in Section \ref{sec:data}.}
    \label{fig:data}
\end{figure*}

\section{Results and Discussion}\label{sec:results}
\subsection{Bar Measurement}\label{ssec:bar}
We integrate our set of observed stars in each of the selected BFE models and measure dynamical lengths for each ILR population. \RILRh, \RILRw, and \RILRc correspond to the dynamical length measured using the Hot, Warm, and Cool ILR stars, respectively. We then chose the 6 ``best fit" models based on the fractional difference between the intrinsic BFE model \RILRw and the \RILRw measured using our observational data. For each of the 6 potentials, we compute an estimate of the distribution of the 3 \RILR's using the Monte-Carlo (MC) method of error propagation. We used 50 samples of input data, drawn from a multivariate Gaussian distribution defined by the covariance matrix of the 6D phase-space coordinates from Section \ref{ssec:data-6d}. We then integrate each sample across each potential model, and compute the frequencies and dynamical length. This gives us a distribution of \RILRh, \RILRw, and \RILRc values for each potential model and we compute the median and the 16th and 84th percentiles to report as the uncertainty.

One of the 6 chosen potential models is, in fact, the potential model that \citetalias{lucey2023} find to be the best fit, \texttt{agama\_00301}. However, in their analysis, the authors fix all pattern speeds to 41 km/s/kpc. To see if we can reproduce the results, we repeat our analysis for \texttt{agama\_00301} using this pattern speed. With the Cartesian frequency cuts, we find an \Rfreq of $\sim9.7$ kpc, which is much larger than the $\sim3.5$ kpc found by the authors. This was due to outliers in the apo-centre distribution, centred around 12 kpc. Once removed, we replicate the 3.5 kpc results. We include the results for the other orbit populations in the subsequent figures as \texttt{agama\_00301\_L23}, but we do not include it as a best fit model candidate.

Upon further inspection, 9 of the 22 potential models tested show significant outliers in the apo-centre distribution for stars classified by the Cartesian frequency cuts. This significantly biases the measurement of the bar to much larger lengths. In contrast, none of the Warm ILR distributions show significant outliers. In part, this is likely due to the fact that, for Warm ILR, we have made a cut in $\Omega_z/\Omega_R$, which is correlated with radial extent. However, there is generally significant overlap in the apo-centre distributions of the Warm and Cool ILR stars, which indicates that we are not simply removing the end of the distribution by making a cut in distance. This can be seen in Figure \ref{fig:apo_dist}, which shows the apo-centre distributions of each orbit population for the N-body samples of \texttt{agama\_00419}.

Figure \ref{fig:length-diff-warm} shows the fractional difference between the intrinsic \RILRw of the potential model and the \RILRw measured using the observational data versus the potential model’s intrinsic \RILRw. The black points represent the fractional length differences for all of the potential models. The coloured points represent the median and confidence intervals for the 6 models with the best fit \RILRw and \texttt{agama\_00301\_L23}, which are connected via a gray band to the original difference measurement used to determine the best-fit models. The marker symbols denote the corresponding potential model. In this view, we see that 4 of the 6 potentials fall below a fractional difference of 5 percent with errors of a few percent -- spanning a range of bar lengths from $\sim2.2$--6 kpc.

\begin{figure}
    \centering
    \includegraphics[width=\linewidth]{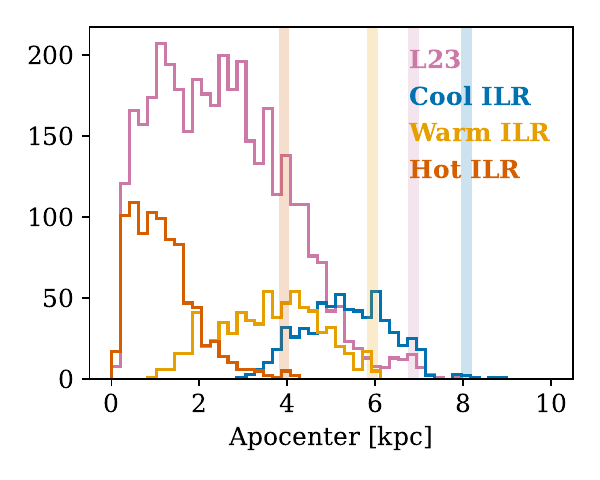}
    \caption{Histogram of apo-centre distances [kpc] for \texttt{agama\_00419} N-body samples. The purple lines represent stars classified using the \citetalias{lucey2023} frequency cuts, the blue are Cool ILR, the orange are Warm ILR, and the red are Hot ILR. The coloured vertical lines show the dynamical length estimate determined from the corresponding resonant orbit population.}
    \label{fig:apo_dist}
\end{figure}

\begin{figure*}[t]
    \centering
    \includegraphics[width=0.9\linewidth]{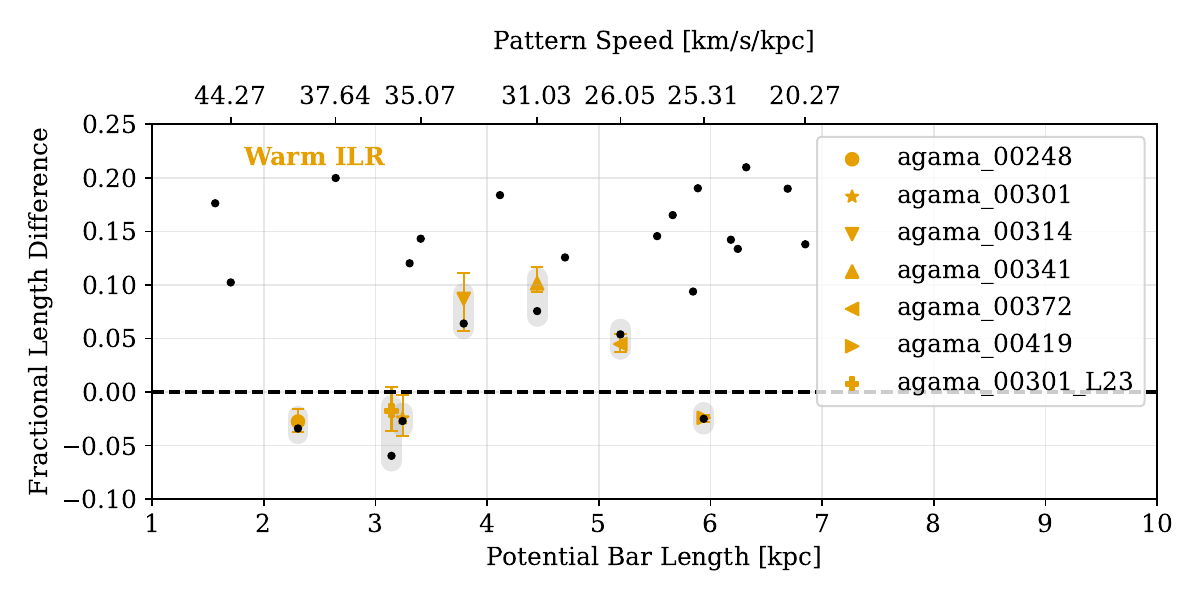}
    \caption[]{The fractional differences between the intrinsic Warm ILR dynamical length of the potential model and that derived from the observational data versus the potential model's intrinsic Warm ILR dynamical length. The top axis represents the pattern speed for that bar length\footnotemark. The black points represent all of the potential models and the orange points represent the median and confidence intervals from Monte-Carlo analysis, which are connected to the corresponding black point by a gray band. The different markers denote the different potential models.}
    \label{fig:length-diff-warm}
\end{figure*}

\begin{figure*}
    \centering
    \includegraphics[width=0.9\linewidth]{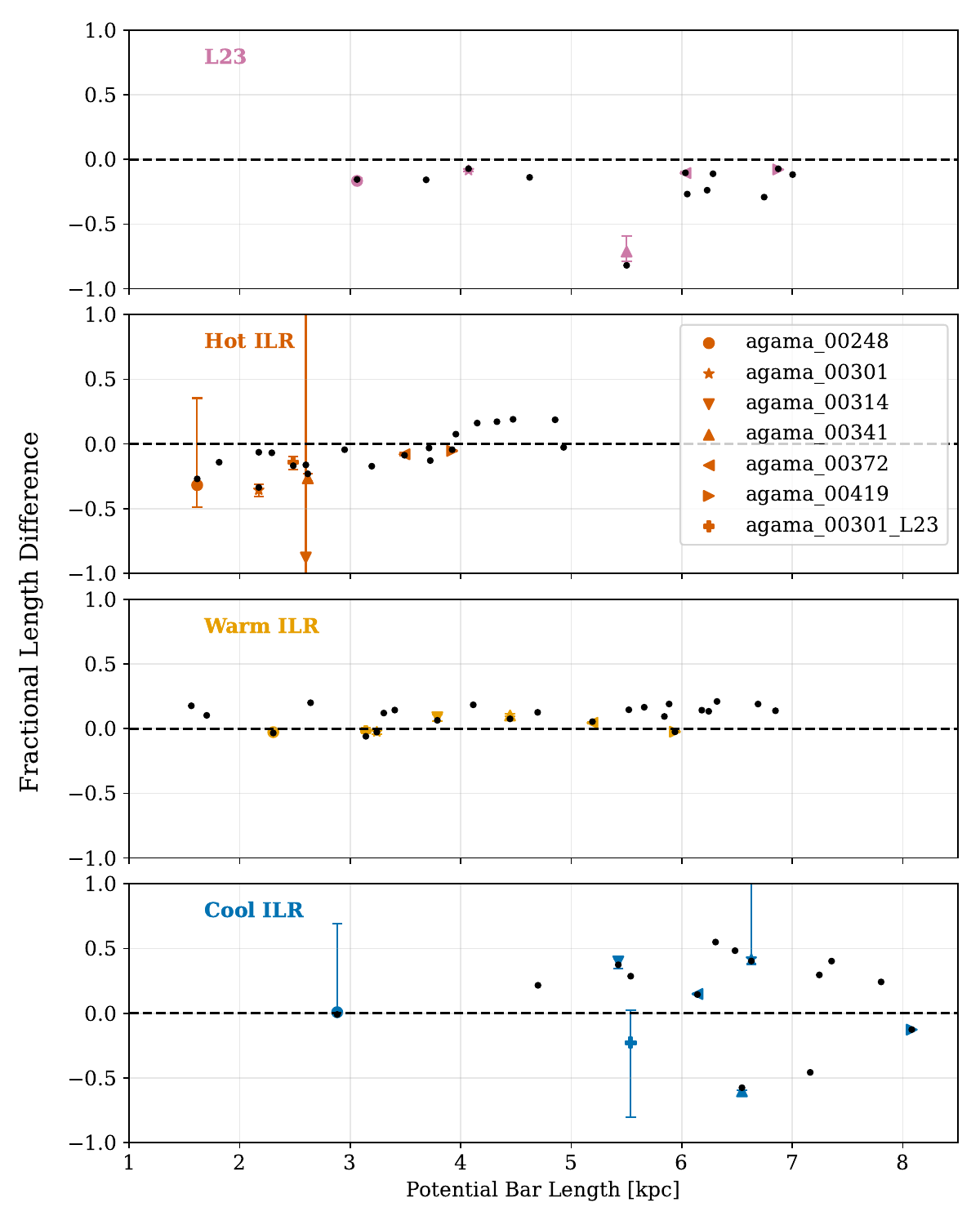}
    \caption{The fractional difference between the intrinsic dynamical length of the potential model and that of the observational data versus the potential model's intrinsic dynamical length. Each panel represents the dynamical lengths for a distinct population of stars: \citetalias{lucey2023} Cartesian cuts (top; purple), Hot ILR (top centre; red), Warm ILR (bottom centre; orange), and Cool ILR (bottom; blue). The black points in each panel represent all of the potential models (some lie outside of the figure). The coloured points represent the median and confidence intervals from the Monte-Carlo analysis for the 6 potentials with the best fit \RILRw and \texttt{agama\_00301\_L23} (best fit from \citetalias{lucey2023} with a pattern speed of 41 km/s/kpc), and the marker symbols (shown in the legend for Hot ILR panel) denote the corresponding potential model. \textit{Note: To preserve readability, we neglect a small number of points that fall outside of the current bounds of the figure.}}
    \label{fig:length-diff}
\end{figure*}

For completeness, Figure \ref{fig:length-diff} shows the fractional difference between the intrinsic dynamical length of the potential model (instrinsic \RILR) and the \RILR derived using the observational data versus the potential model’s intrinsic \RILR for each orbit population. Each panel is on the same axis scaling and represents a distinct population: \citetalias{lucey2023} Cartesian cuts (top; purple), Hot ILR (top centre; red), Warm ILR (bottom centre; orange), and Cool ILR (bottom; blue). The coloured points represent the median and confidence intervals for the 6 models with the best fit \RILRw and \texttt{agama\_00301\_L23}, and the marker symbols (shown in the legend in the Hot ILR panel) denote the corresponding potential model. The black points in each panel represent the remaining potential models (some lie outside of the figure). This figure illustrates that, of the ILR populations, only the Warm ILR measurements are free of large changes in length difference due to variations in the initial conditions of the orbit integration. In addition, we see that the Cool ILR stars are overall biased toward both larger lengths \textit{and} larger length differences. The Warm and Hot ILR orbits have similar fractional differences on average, but the warm ILR population extends to larger lengths, as expected. 

We note that the orbit populations derived from the \citetalias{lucey2023} frequency cuts suggest larger bar lengths on average than the Warm ILR populations, and often larger intrinsic \Rfreq's for the same potential models than was found by \citetalias{lucey2023}. A likely explanation for this difference is that we use the pattern speed, as computed from the N-body model, where as \citetalias{lucey2023} use 41 km/s/kpc for every potential model. This would bring in the corotation radius and likely the length of the bar, which seems to be the case, at least for the author's best-fit model \texttt{agama\_00301}. For this potential, we measure an \Rfreq of 4 kpc using the intrinsic pattern speed of 35.9 km/s/kpc, and an \Rfreq of 3.5 kpc using the 41 km/s/kpc from \citetalias{lucey2023}. In this case, increasing the pattern speed does seem to have decreased the measured bar length using the Cartesian frequency cuts. This is mirrored, to a lesser extent, by the \RILR measurements (see Figure \ref{fig:length-diff}).

The point placements and uncertainties in Figures \ref{fig:length-diff-warm} and \ref{fig:length-diff} may be partially understood in light of the snapshot morphology. Figure \ref{fig:snapshots} shows the face-on, stellar surface density of the corresponding N-body snapshot for each of the 6 ``best-fit" models. The orange and blue bars represent the location of the potential's intrinsic \RILRw and the measured \RILRw of the data, respectively. This figure shows that, at least qualitatively, we recover reasonable \RILRw values based on the density of the underlying potential. The exception is \texttt{agama\_00248}, where the bar is clearly still forming and does not have a distinct shape. This is consistent with the fact that its fractional errors in \RILRh and \RILRc are very large. In addition, we can see that models with the best fitting Cool ILR lengths are those with minimal asymmetries towards the outer edges of the bar and beyond (e.g., \texttt{agama\_00419}). We can spot a similar trend in the Hot ILR -- those with larger fractional differences have noticeable warping within the bar region (e.g., \texttt{agama\_00301}).

\begin{figure*}
    \centering
    \includegraphics[width=0.9\linewidth]{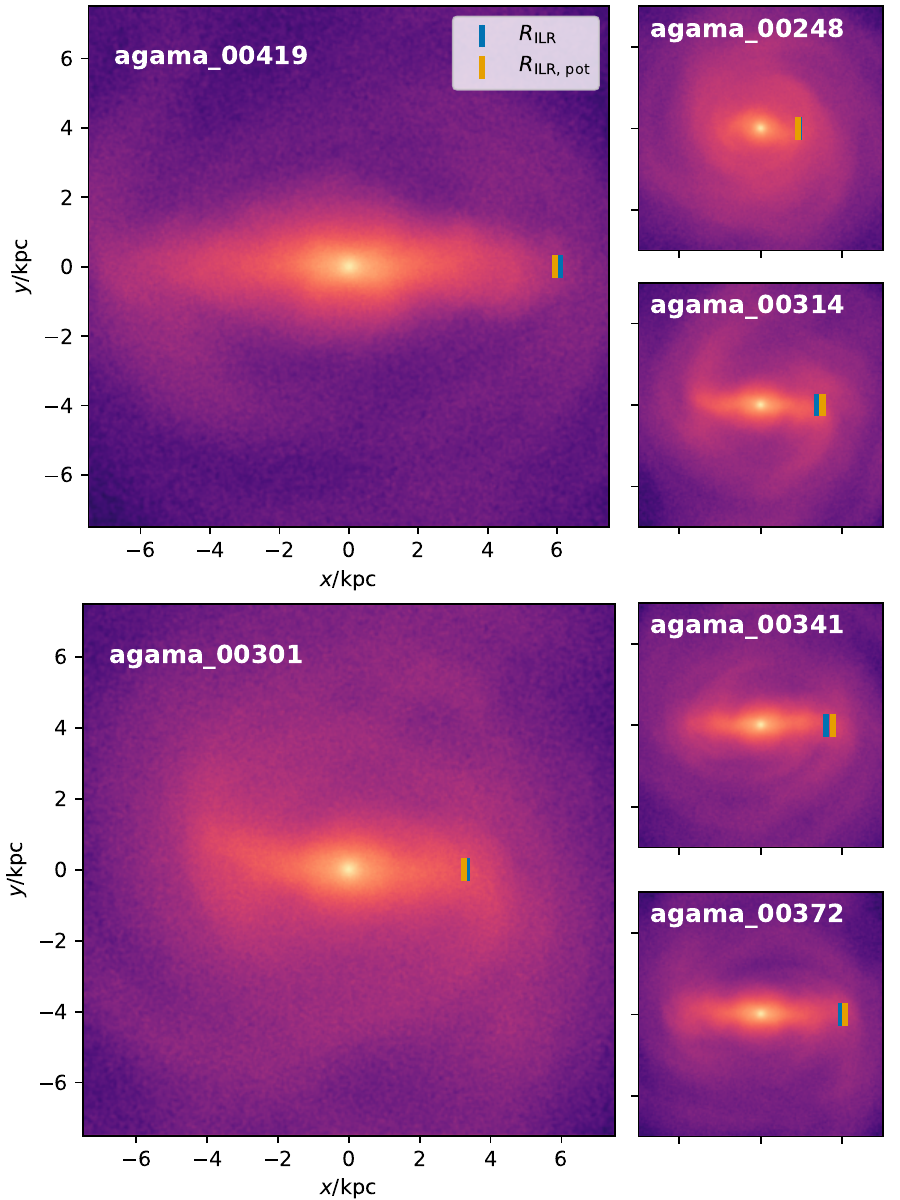}
    \caption{Face-on, stellar surface density plot for the 6 ``best-fit" snapshots ($\S$ \ref{sec:results}). The blue and orange line-segments represent the \RILR measured using the observational data, and the intrinsic potential \RILR, respectively. Error bars are omitted as the largest standard deviation is just $\sim$100 pc. The annotated ``agama\_00XXX'' is the file name for the potential model, where the ascending numbers correspond to increasing simulation time. The colour-map in each panel is set to the same scale.}
    \label{fig:snapshots}
\end{figure*}

In summary, these results show that 4 potential models produce consistent \RILRw's to within 5 percent. However, these results do not clearly distinguish one potential model as \textit{the} best fit, as the disparity between fractional \RILRw differences is at most $\sim1$ percent. If we remove \texttt{agama\_00248}, which does not have a visually distinguishable bar, we are left with \texttt{agama\_00301}, \texttt{agama\_00372}, and \texttt{agama\_00419}, which have intrinsic \RILRw's of 3.24, 5.19, and 5.94 kpc, respectively, and pattern speeds of 35.9, 26.05, and 25.3 km/s/kpc, respectively.

\subsection{Frequency Analysis}\label{ssec:results_freqs}
We now turn to examining the frequency space of our best fit models. For this section, we look at \texttt{agama\_00419}, which has the longest bar, and \texttt{agama\_00301}, which has the shortest bar, in detail. We show the figures for the rest of the 4 models in Appendix \ref{app:freqs}. 

Figures \ref{fig:freq_cuts_data_00419} and \ref{fig:freq_cuts_data_00301} have the same panel layout as Figure \ref{fig:freq_cuts_samples} for the \texttt{agama\_00419} and \texttt{agama\_00301} models, respectively. The points now represent our survey data rather than the N-body samples. 

For \texttt{agama\_00419}, we see a qualitatively similar structure in the frequency space to that of Figure \ref{fig:freq_cuts_samples}; the cloud of stars surrounding the ILR around $0.5<\Omega_z/\Omega_R<1.25$; the clustering near $0.5 < \Omega_z/\Omega_R < 1.0$ and $-1.0 < (\Omega_\phi - \Omega_p) / \Omega_R < -0.5$; the ``bump" in the ILR population near $\Omega_z/\Omega_R = 1.2$, which appears near the intersection of the ILR and the ``vertical" ILR ($\left(\Omega_\phi - \Omega_p\right) / \Omega_z = 1/2$; discussed in \citetalias{beraldoesilva2023}); and the stars on the $3\Omega_x - \Omega_y = 0$ resonance, which have orbits similar to the bifurcated x1 orbits (e.g., row 5 in Figure 4 of \citet{valluri2016} or the x1b panel of Figure 5 in \citet{Petersen2021}).

\begin{figure*}
    \centering
    \includegraphics[width=\linewidth]{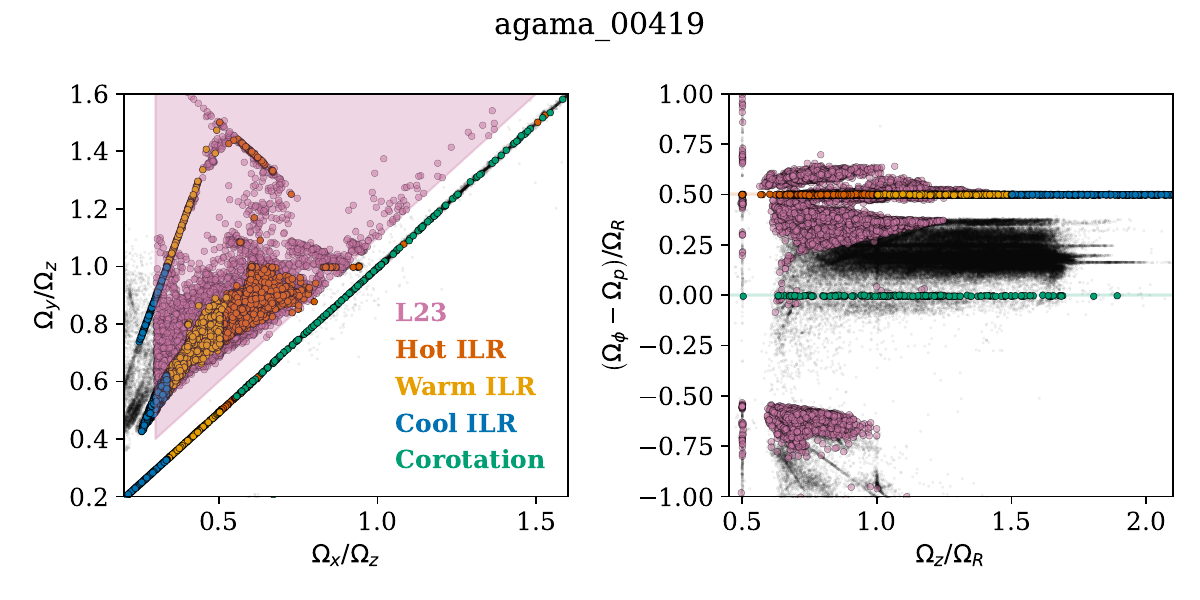}
    \caption{Fundamental frequencies for orbits of observed stars in the \texttt{agama\_00419} potential (an analog of Figure \ref{fig:freq_cuts_samples}). The red points represent the hot ILR, the orange are warm ILR, the blue are cool ILR, the green are corotation resonance stars, and the black points are any stars not classified into any of the preceding populations. Left panel: Cartesian frequency ratios in the bar-rotating frame. The purple shaded region is the cut used in \citetalias{lucey2023}. Right panel: Cylindrical coordinate frequencies in the inertial frame. The purple points are those that lie in the shaded region in the left panel. The purple bar cut misses many ILR stars while including non-ILR stars.}
    \label{fig:freq_cuts_data_00419}
\end{figure*}

\begin{figure*}
    \centering
    \includegraphics[width=\linewidth]{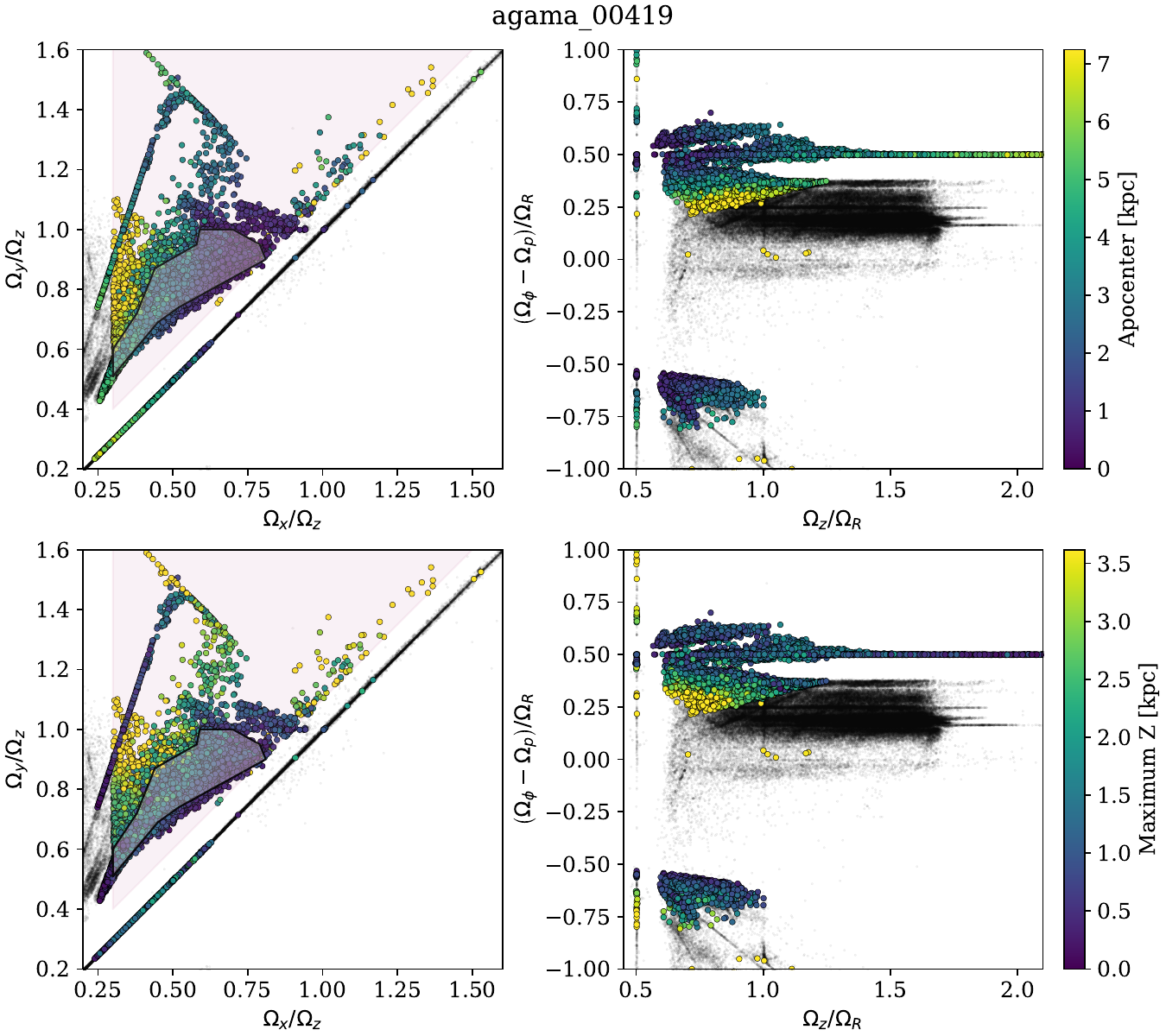}
    \caption{Fundamental frequencies for orbits of observed stars in the \texttt{agama\_00419} potential (same stars as in Figure \ref{fig:freq_cuts_data_00419}; analogue of Figure \ref{fig:freq_z_apo_samples}). The left and right panels are the same as in Figure \ref{fig:freq_cuts_data_00419}, the colours now represent (top) the apo-centre distance and (bottom) the maximum z extent along the orbit for the \citetalias{lucey2023} cuts and the Hot, Warm, and Cool ILR stars. The outlined region in the left panels are a visual aid to show roughly where the bulk of ILR stars overlap with the Cartesian frequency cuts in Figure \ref{fig:freq_cuts_data_00419}. We see that choosing ILR stars to measure the dynamical length removes a region of large apo-centre distances.}
    \label{fig:freq_z_apo_data_00419}
\end{figure*}

Similarly, the \texttt{agama\_00301} frequency space for our data set replicates much of the same structure seen in the N-body samples (see Figure \ref{fig:freq_cuts_00301}). In this case, we note a dearth of Cool ILR stars in the data frequency space compared to that of the N-body samples. One explanation is that this is due to the presence of spiral arms in the N-body potential. Since the Cool ILR stars are more disky and extend the furthest in apo-centre distance, the stars in data that are near this outer-bar region may be better described as being members of the spiral arms. This same frequency space phenomenon can be seen in the other potentials with prominent asymmetry near the edge of the bar: \texttt{agama\_00314} (Figure \ref{fig:freq_cuts_00314}) and \texttt{agama\_00341} (Figure \ref{fig:freq_cuts_00372}). The most dramatic example is in the top panel in Figure \ref{fig:freq_cuts_00314}, which shows the data stars begin to deviate from $(\Omega_\phi - \Omega_p)/\Omega_R = 0.5$ shortly after the transition from Warm to Cool ILR. 

\begin{figure*}
    \centering
    \includegraphics[width=\linewidth]{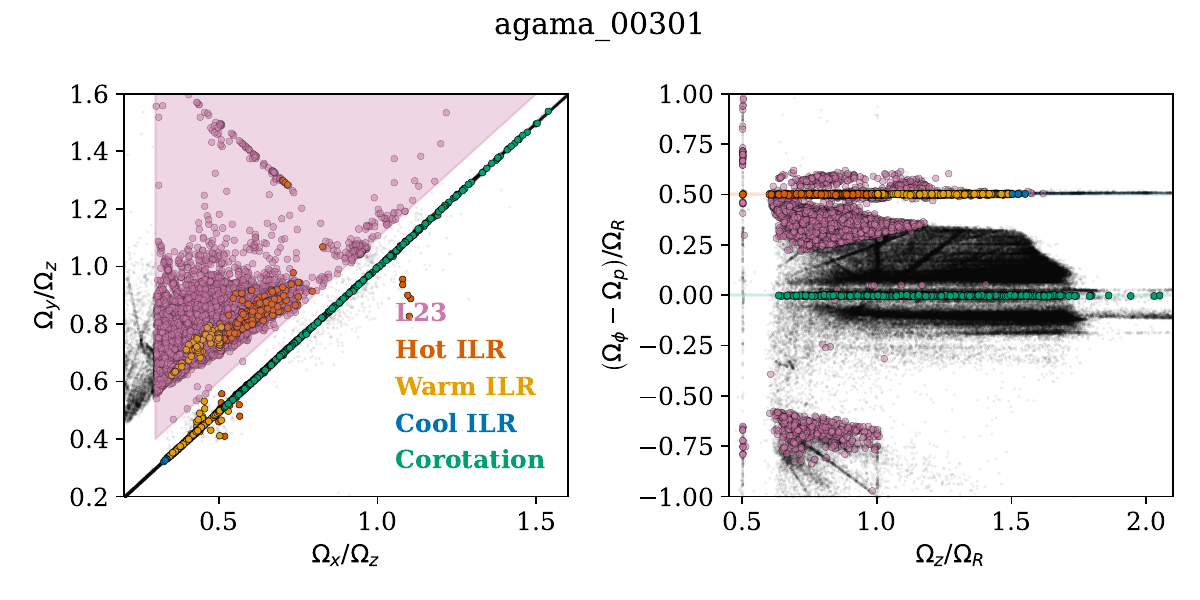}
    \caption{Fundamental frequencies for orbits of observed stars in the \texttt{agama\_00301} potential (an analogue of Figure \ref{fig:freq_cuts_samples}). The red points represent the hot ILR, the orange are warm ILR, the blue are cool ILR, the green are corotation resonance stars, and the black points are any stars not classified into any of the preceding populations. Left panel: Cartesian frequency ratios in the bar-rotating frame. Purple shaded region is the cut used in \citetalias{lucey2023}. Right panel: Cylindrical coordinate frequencies in the inertial frame. The purple bar cut misses many ILR stars while including non-ILR stars.}
    \label{fig:freq_cuts_data_00301}
\end{figure*}

\begin{figure*}
    \centering
    \includegraphics[width=\linewidth]{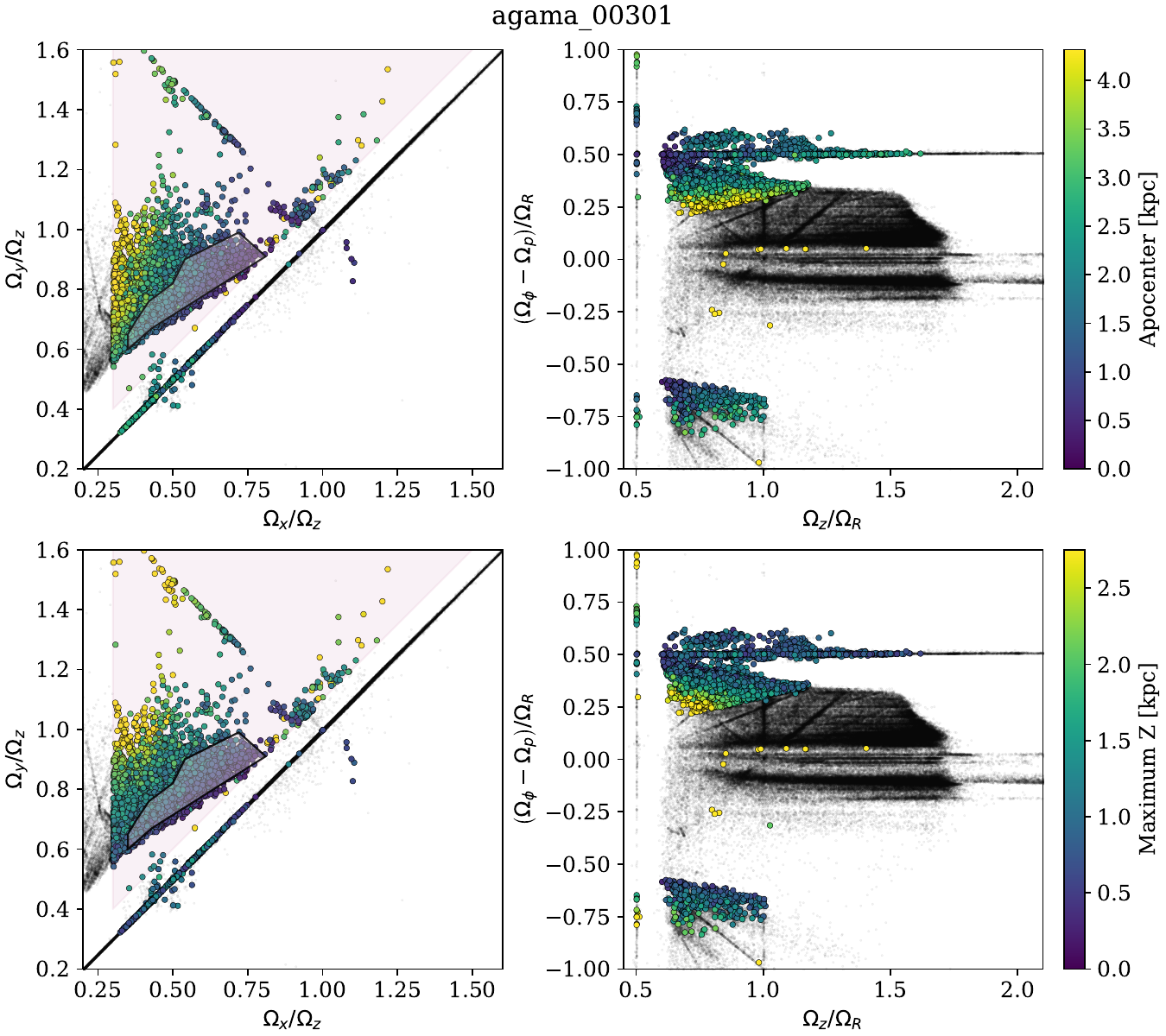}
    \caption{Fundamental frequencies for orbits of observed stars in the \texttt{agama\_00301} potential (same stars as in Figure \ref{fig:freq_cuts_data_00301}; analogue of Figure \ref{fig:freq_z_apo_samples}). The left and right panels are the same as in Figure \ref{fig:freq_cuts_data_00301}, the colours now represent (top) the apo-centre distance and (bottom) the maximum z extent along the orbit for the \citetalias{lucey2023} cuts and the Hot, Warm, and Cool ILR stars. The outlined region in the left panels are a visual aid to show roughly where the bulk of ILR stars overlap with the Cartesian frequency cuts in Figure \ref{fig:freq_cuts_data_00301}. We see that choosing ILR stars to measure the dynamical length removes a region of large apo-centre distances.}
    \label{fig:freq_z_apo_data_00301}
\end{figure*}

Figures \ref{fig:freq_z_apo_data_00301} and \ref{fig:freq_z_apo_data_00419} are the data analogues of Figure \ref{fig:freq_z_apo_samples} for \texttt{agama\_00301} and \texttt{agama\_00419}. This again shows that the cylindrical frequency cuts remove a large amount of high apo-centre stars, and recover many more ILR stars. 

\footnotetext{The dynamical length roughly increases as the pattern speed decreases. However, the individual measurements will not be strictly increasing with pattern speed. For the purposes of the figure, we simply pick a bar length measurement and make the tick mark the corresponding pattern speed value to show the rough trend and range of values.}

\section{Conclusion} \label{sec:con}
In this paper, we present an updated frequency analysis method for measuring the dynamical length of a galactic bar. We build on the methods of \cite{lucey2023} by using the updated numerical methods and bar orbit classification criteria from \cite{beraldoesilva2023}. These updates allow us to isolate the looped x1, or ``Warm" ILR, orbits that support the shoulder regions of the bar. We show that this updated criterion captures many of the bar supporting orbits that were previously missed while excluding those that are still resonant orbits, but clearly do not contribute to the morphology of the bar (e.g., Figure \ref{fig:orbits}). In addition, this cut in frequency space inherently removes outliers in the apo-centre distance (e.g., Figure \ref{fig:freq_z_apo_samples}), which means that we can reliably measure the extent of the distribution.

We applied these methods to the MW by measuring the dynamical length using stars in the Gaia, APOGEE, and OLGE data sets. This allows us to determine a set of 3 dynamically consistent potential models that produce self-consistent bar lengths to within 5 percent (excluding \texttt{agama\_00248}; see Section \ref{sec:results}). Table \ref{tab:final_pots} shows the properties of these potential models; the name of the potential; the pattern speed in km/s/kpc; and the \RILRw as measured from the N-body samples (the ``intrinsic" dynamical length of the potential model).  

\begin{table}
\centering
\caption{Properties of the 3 most dynamically consistent potential models. The columns are the name of the potential file, the pattern speed in km/s/kpc, and the intrinsic \RILRw in kpc.}
\begin{tabular}{ccc}
    \hline
    Potential & Pattern Speed & \RILRw \\
              & km/s/kpc & kpc \\
    \hline
    \texttt{agama\_00301} & 35.901 & 3.242 \\
    \texttt{agama\_00372} & 26.051 & 5.193 \\
    \texttt{agama\_00419} & 25.305 & 5.94 \\
    \hline
\end{tabular}
\label{tab:final_pots}
\end{table}

From our selection of N-body snapshots, we determine that \texttt{agama\_00419}, \texttt{agama\_00372}, and \texttt{agama\_00301} best describe our data set. These potential models have intrinsic, Warm ILR dynamical lengths that are consistent with the data to within 5 percent. The first 2 of these models have much larger and slower bars than most measurements found by other methods \citep[e.g.,][]{Saito2011,Wegg2013,Wegg2015,Zhang2024,Vislosky2024,hunt2025}, with the exception of \cite{horta2024}. The third model, \texttt{agama\_00301}, is the same potential that \citetalias{lucey2023} found to be the most consistent, and is more in line with most measurements of the length and pattern speed. We do not apply any prior information about other measurements in our modelling.

For our potential models, we chose snapshots from a single N-body simulation of a MW-like galaxy. This provides us with a dynamically consistent set of barred potentials with a bar that has formed and evolved in a realistic way -- coupling the mass distribution and pattern speed to set the dynamical length. Choosing a N-body model also allows us to draw self-consistent samples from the underlying distribution function, which is required in order to measure intrinsic dynamical lengths. However, the underlying distribution function for any given snapshot is not guaranteed to well describe the MW, or be at all similar to the MW today. Consistent dynamical lengths only tell us that a given snapshot, pattern speed, and set of observed stars produces similar orbits to those intrinsic to the N-body snapshot. Exploring a wider range of potential models will give us insight into the degeneracies of this parameter space.

In addition, we have made a number of other choices and assumptions that leave us with directions for future work:

\begin{enumerate} 
    \item For every model, we have assumed a bar angle of 27 degrees -- the same as \citetalias{lucey2023}. Future work will explore how sensitive this dynamical framework is to the choice of bar angle.

    \item As expected, we find similar, but not identical, orbit morphologies in each subpopulation to that of \citetalias{beraldoesilva2023}. These differences are likely due to the asymmetries in our potential models. More work needs to be done to determine how the choice of frequency boundaries depends on the specific potential model and how much this affects the estimates of the bar length.

    \item Due to computational constraints, we only compute data \RILR's for a subset of the available potential models. Analysing additional potential models may tell us why we seem to prefer longer bar lengths, why the Cool ILR stars are absent for some potential models, and may provide a potential model with even better agreement.

    \item The computation of Monte-Carlo uncertainties is also expensive. In future work, we plan to apply an automatic differentiation \citep{baydin2018} based modelling scheme to propagate errors from the data through our model to the dynamical bar length.
\end{enumerate}

We have tried to make our work as accessible as possible so that others in the community can reproduce our results, and reuse and adapt our methods for their own work. To this end, all of the Python code we used to prepare this paper resides in a GitHub repository\footnote{\url{https://github.com/langfzac/MW_bar_paper}} and the N-body simulation snapshots are publicly available\footnote{\url{https://users.flatironinstitute.org/~jhunt/Bennett_MWP14-3_MR_snapshots/}}. All of the observational data we used is already public, and the instructions for gathering and the code for processing the data can be found in the GitHub repository.

\section*{Acknowledgements}

ZL thanks Sioree Ansar, Carrie Filion, Francesca Fragkoudi, Sarah Pearson, Michael Petersen, and Adrian Price-Whelan for useful discussions that have greatly improved this work. ZL also thanks the anonymous referee for comments that helped to improve this manuscript. ZL and RES were supported by NSF grant AST-2007232, Sloan Foundation award FG-2023-20669, and Simons Foundation grant 1018462 in the completion of this work. ML acknowledges the support of the National Science Foundation under Award No. 2303831. JH acknowledges the support of a UKRI Ernest Rutherford Fellowship ST/Z510245/1. ZL acknowledges the use of the University of Pennsylvania General Purpose Cluster, on which the computations in this paper were carried out.

%%%%%%%%%%%%%%%%%%%%%%%%%%%%%%%%%%%%%%%%%%%%%%%%%%
\section*{Data Availability}

The N-body simulation snapshots are available at \url{https://users.flatironinstitute.org/~jhunt/Bennett_MWP14-3_MR_snapshots/}. 

The code used to gather, process, and analyse the public data sets and generate the figures is available at \url{https://github.com/langfzac/MW_bar_paper}.



\bibliographystyle{mnras}
\bibliography{main} % if your bibtex file is called example.bib

%\begin{thebibliography}{99}
%\end{thebibliography}

%%%%%%%%%%%%%%%%%%%%%%%%%%%%%%%%%%%%%%%%%%%%%%%%%%

%%%%%%%%%%%%%%%%% APPENDICES %%%%%%%%%%%%%%%%%%%%%

\appendix
\section{Frequency Maps}\label{app:freqs}
Figures \ref{fig:freq_cuts_00248}, \ref{fig:freq_cuts_00301}, \ref{fig:freq_cuts_00314}, \ref{fig:freq_cuts_00341}, and \ref{fig:freq_cuts_00419} show frequency maps for the 6 best-fit potential models. Each figure has the same layout and colour scheme. The left panels show the Cartesian frequency space, with the shaded region representing the \citetalias{lucey2023} cuts, and the right panels are the cylindrical frequency space for the same stars. The top panels represent the 10,000 N-body samples and the bottom panels represent the Gaia/APOGEE/OGLE data set for the same potential. The purple points represent stars classified by the \citetalias{lucey2023} cuts, the red are Hot ILR, the orange are warm ILR, blue are Cool ILR, and green are corotation resonant stars.  

\begin{figure*}
    \includegraphics[width=\textwidth]{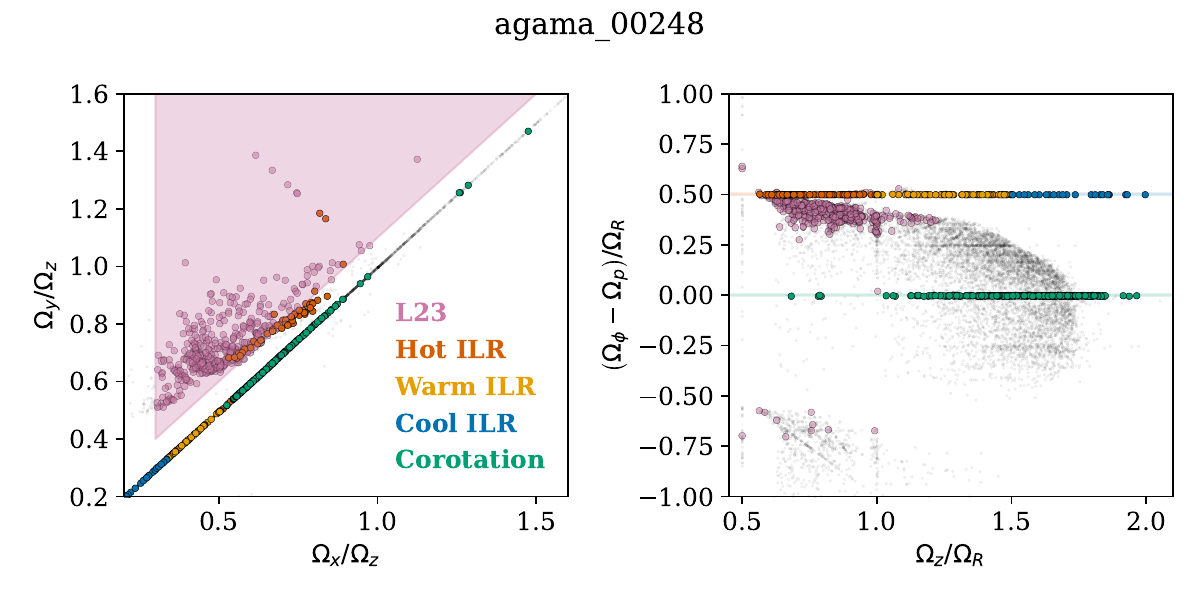}
    \hfill
    \includegraphics[width=\textwidth]{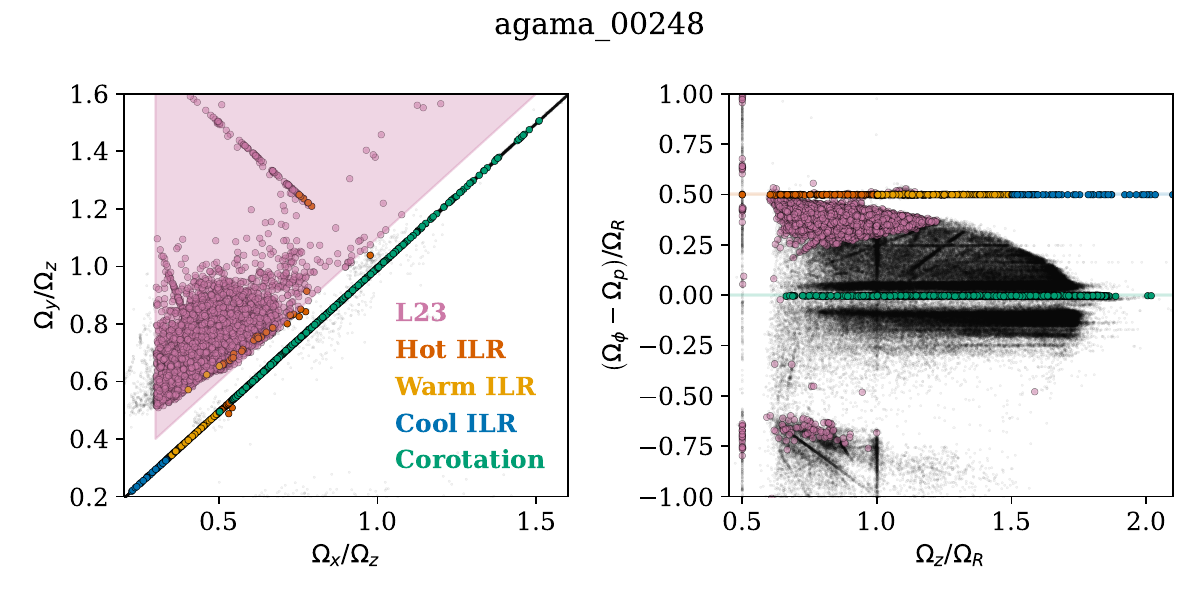}
    \caption{Frequency space for \texttt{agama\_00248}.}
    \label{fig:freq_cuts_00248}
\end{figure*}

\begin{figure*}
    \includegraphics[width=\textwidth]{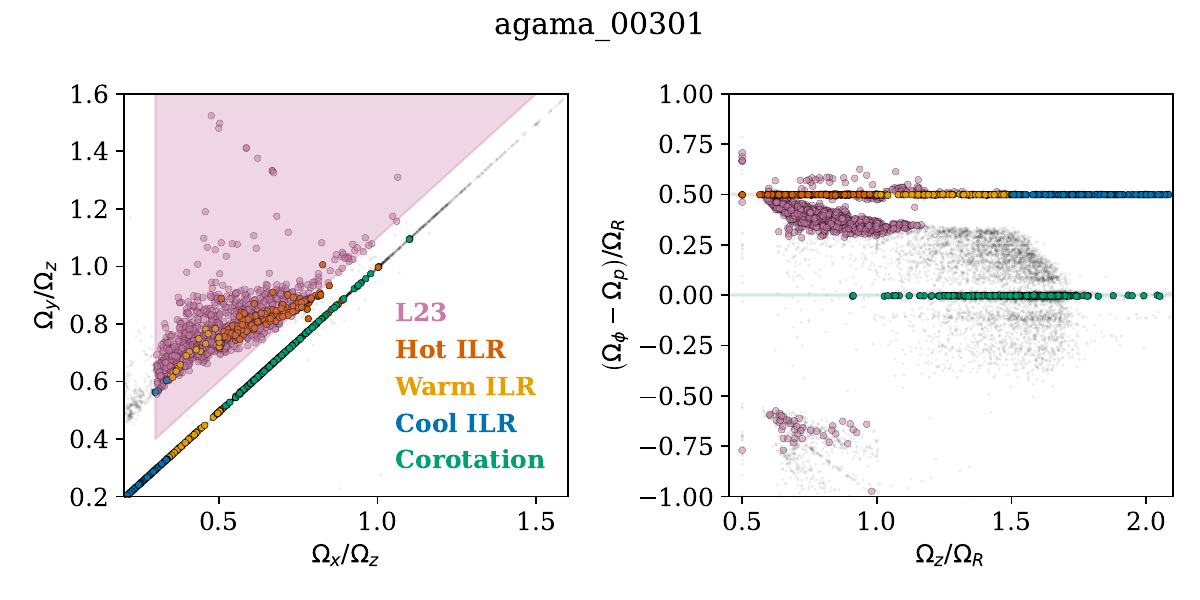}
    \hfill
    \includegraphics[width=\textwidth]{freq_cuts_data_agama_00301.pdf}
    \caption{Frequency space for \texttt{agama\_00301}.}
    \label{fig:freq_cuts_00301}
\end{figure*}

\begin{figure*}
    \includegraphics[width=\textwidth]{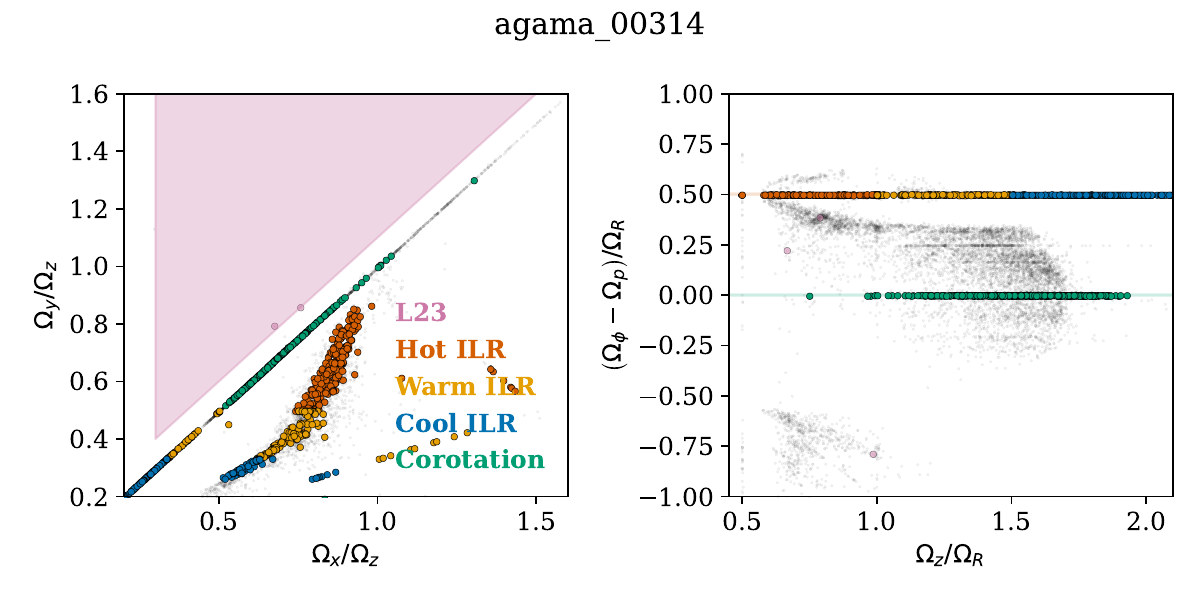}
    \hfill
    \includegraphics[width=\textwidth]{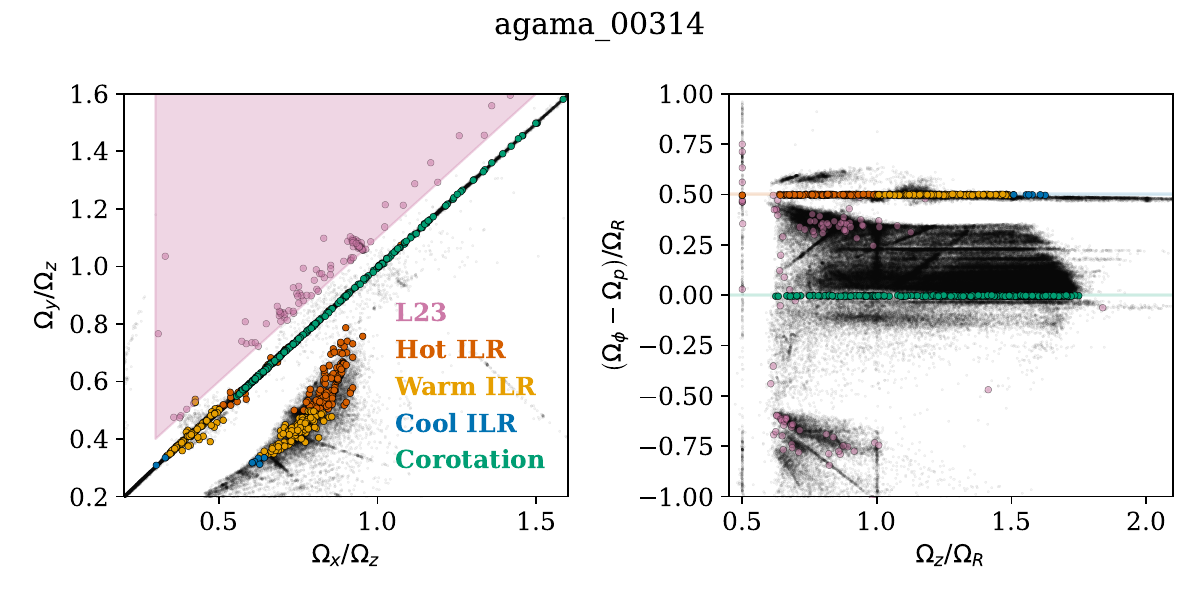}
    \caption{Frequency space for \texttt{agama\_00314}.}
    \label{fig:freq_cuts_00314}
\end{figure*}

\begin{figure*}
    \includegraphics[width=\textwidth]{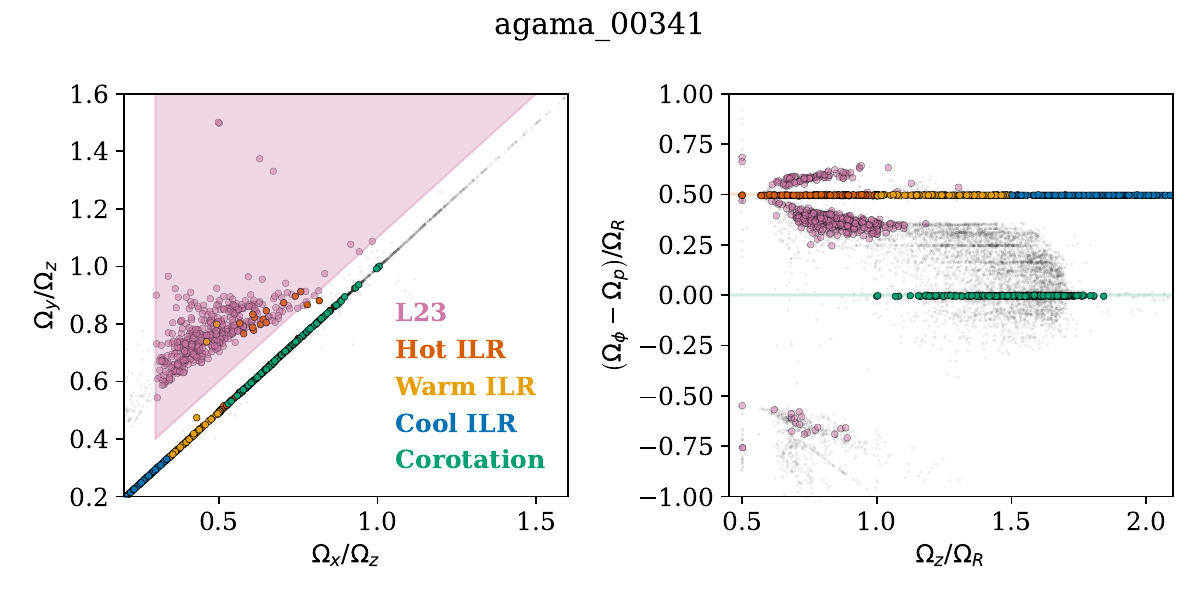}
    \hfill
    \includegraphics[width=\textwidth]{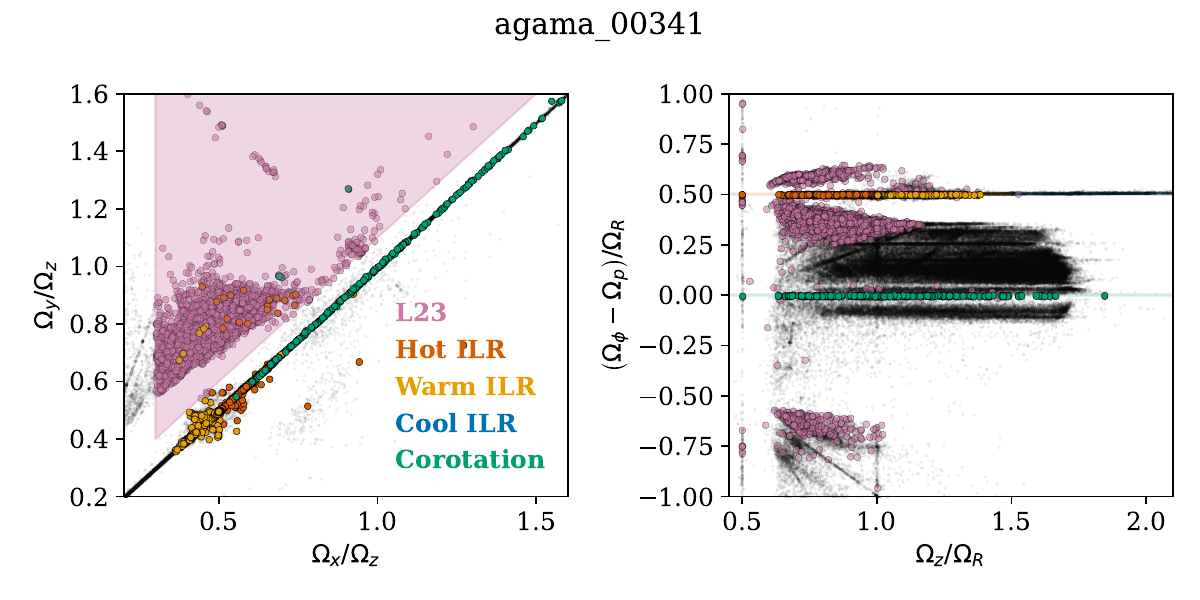}
    \caption{Frequency space for \texttt{agama\_00341}.}
    \label{fig:freq_cuts_00341}
\end{figure*}

\begin{figure*}
    \includegraphics[width=\textwidth]{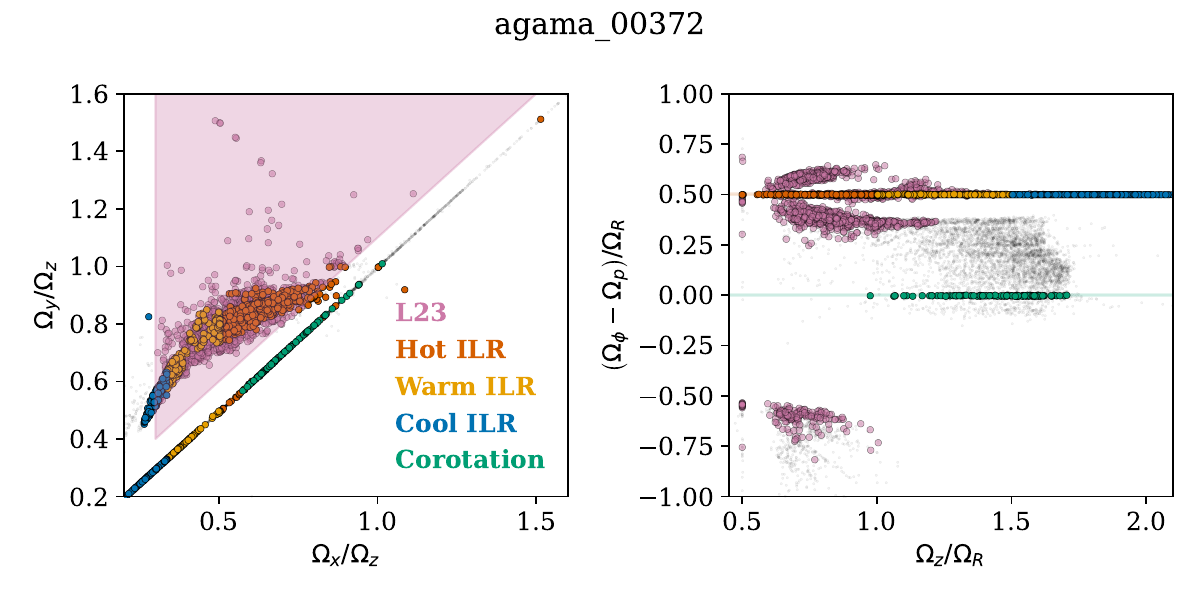}
    \hfill
    \includegraphics[width=\textwidth]{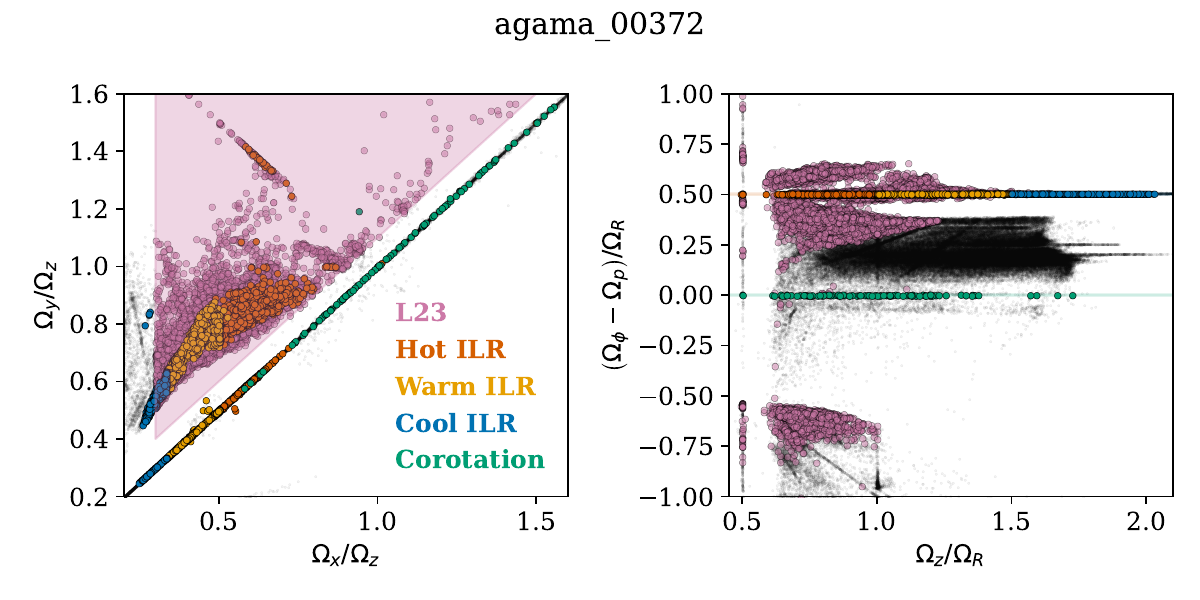}
    \caption{Frequency space for \texttt{agama\_00372}.}
    \label{fig:freq_cuts_00372}
\end{figure*}

\begin{figure*}
    \includegraphics[width=\textwidth]{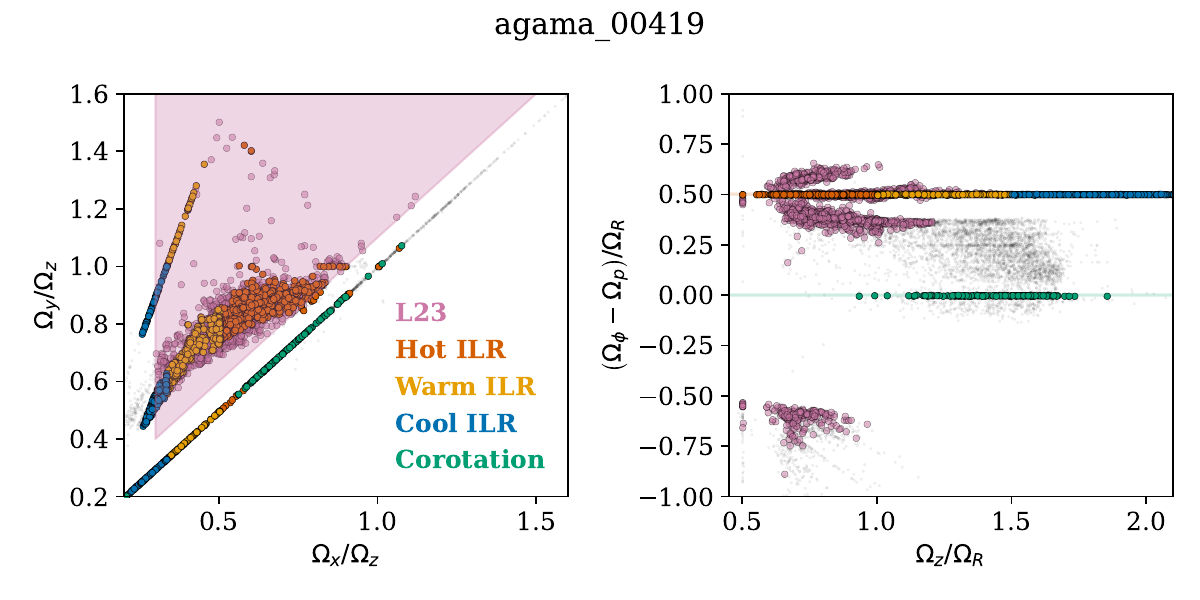}
    \hfill
    \includegraphics[width=\textwidth]{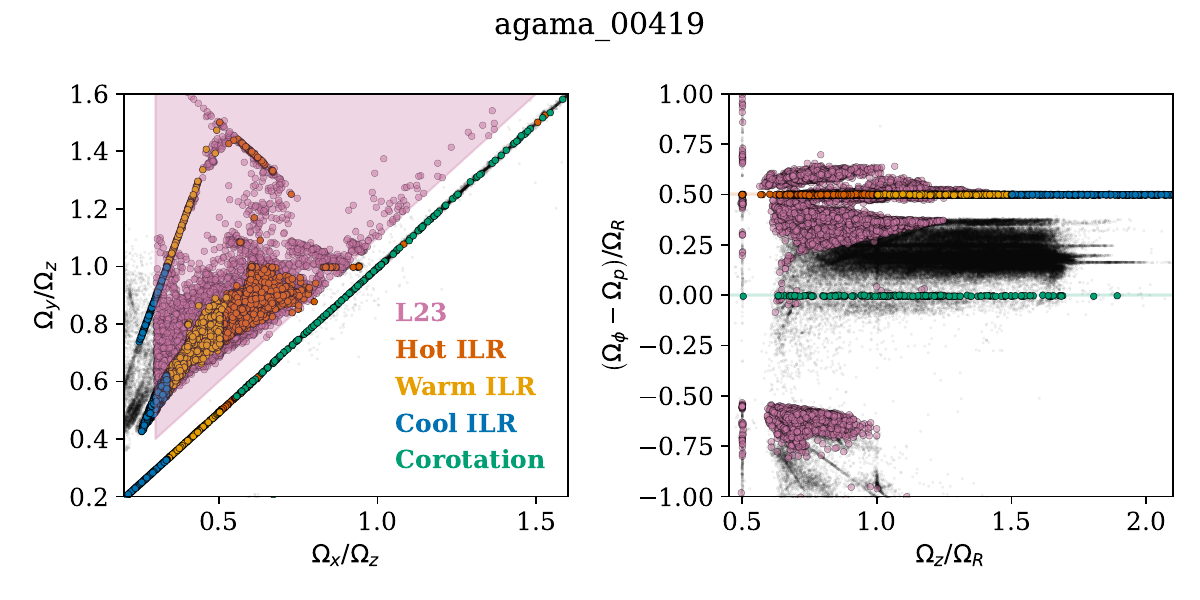}
    \caption{Frequency space for \texttt{agama\_00419}.}
    \label{fig:freq_cuts_00419}
\end{figure*}
%%%%%%%%%%%%%%%%%%%%%%%%%%%%%%%%%%%%%%%%%%%%%%%%%%

% Don't change these lines
\bsp	% typesetting comment
\label{lastpage}
\end{document}